\documentclass[acmlarge]{acmart}

\usepackage{pifont}
\usepackage{makecell}
\newcommand{\cmark}{\ding{51}}%
\newcommand{\xmark}{\ding{55}}%
\usepackage{enumerate}
\usepackage{url}
\usepackage{multirow}
\usepackage{tikz}
\usepackage{adjustbox}
\usepackage{graphics}
\usepackage{nicematrix}
\usepackage{xcolor}
\usepackage{enumitem}

\AtBeginDocument{%
  \providecommand\BibTeX{{%
    \normalfont B\kern-0.5em{\scshape i\kern-0.25em b}\kern-0.8em\TeX}}}

\setcopyright{acmcopyright}
\copyrightyear{2023}
\acmYear{2023}
\acmDOI{10.1145/nnnnnnn.nnnnnnn}

\setcitestyle{numbers}

\begin{document}

\title{{Explainable Deep Learning Methods in Medical Image Classification: A Survey}}

\author{Cristiano Patrício}
\email{cristiano.patricio@ubi.pt}
\author{João C. Neves}
\email{jcneves@di.ubi.pt}
\affiliation{%
  \institution{University of Beira Interior and NOVA LINCS}
  \city{Covilhã}
  \country{Portugal}
  \postcode{6201-001}
}
\author{Luís F. Teixeira}
\email{luisft@fe.up.pt}
\affiliation{%
  \institution{University of Porto and INESC TEC}
  \city{Porto}
  \country{Portugal}
  \postcode{4200-465}
}

\renewcommand{\shortauthors}{Patrício et al}

\begin{abstract}

The remarkable success of deep learning has prompted interest in its application to medical imaging diagnosis. Even though state-of-the-art deep learning models have achieved human-level accuracy on the classification of different types of medical data, these models are hardly adopted in clinical workflows, mainly due to their lack of interpretability. The black-box-ness of deep learning models has raised the need for devising strategies to explain the decision process of these models, leading to the creation of the topic of eXplainable Artificial Intelligence (XAI). In this context, we provide a thorough survey of XAI applied to medical imaging diagnosis, including visual, textual, example-based and concept-based explanation methods. Moreover, this work reviews the existing medical imaging datasets and the existing metrics for evaluating the quality of the explanations. In addition, we include a performance comparison among a set of report generation-based methods. Finally, the major challenges in applying XAI to medical imaging and the future research directions on the topic are also discussed.
\end{abstract}



\maketitle

\section{Introduction}
\label{sec:introduction}








%

The progress made on the last decade in the field of artificial intelligence (AI) has supported a dramatic increase in the accuracy of most computer vision applications. Medical image analysis is one of the applications where the progress made assured human-level accuracy on the classification of different types of medical data (e.g., chest X-rays \cite{majkowska2020chest}, corneal images \cite{xu2021deep}). However, and in spite of these advances, automated medical imaging is seldom adopted in clinical practice. According to Zachary Lipton~\cite{lipton2017doctor}, the explanation to this apparent paradox is straightforward, doctors will never trust the decision of an algorithm without understanding its decision process. This fact has raised the need for producing strategies capable of explaining the decision process of AI algorithms, leading subsequently to the creation of a novel research topic named as eXplainable Artificial Intelligence (XAI). According to DARPA~\cite{gunning2019darpa}, XAI aims to ``\textit{produce more explainable models, while maintaining a high level of learning performance (prediction accuracy); and enable human users to understand, appropriately, trust, and effectively manage the emerging generation of artificially intelligent partners}''. In spite of its general applicability, XAI is particularly important in high-stake decisions, such as clinical workflow, where the consequences of a wrong decision could lead to human deaths. This is also evidenced by European Union’s General Data Protection Regulation (GDPR) law, which requires an explanation of the decision-making process of the algorithm, thus improving its transparency before it can be used for patientcare~\cite{goodman2017european}. 

Considering this, it is of utmost importance to invest in the research of novel strategies to improve the interpretability of deep learning methods before being possible to deploy them into clinical practice. During the last years, the research on this topic has focused primarily on devising methods for indirectly analysing the decision process of pre-built models. These methods operate either by analysing the impact of specific regions of the input images on the final prediction (perturbation-based methods~\cite{ribeiro2016should, lundberg2017unified} and occlusion-based methods~\cite{zeiler2014visualizing}) or inspecting the network activations (saliency methods~\cite{selvaraju2017grad, zhou2016learning}). The fact that these methods can be applied to arbitrary network architectures without requiring an additional customization of the model has supported their popularity in the early days of XAI. However, it has been recently shown that post-hoc strategies suffer from several drawbacks regarding the significance of the explanations~\cite{rudin2019stop, adebayo2018sanity}. As a consequence, researchers have focused their attention in the design of models/architectures capable of explaining their decision process \textit{per se}. Inherently interpretable models are believed to be particularly useful in medical imaging~\cite{rudin2019stop}, justifying the recent growth in the number of medical imaging works focusing on this paradigm rather the traditional post-hoc strategies~\cite{kim2021xprotonet,wickramanayake2021comprehensible}. In spite of the recent popularity of inherently interpretable models, the existing surveys on the interpretability of deep learning applied to medical imaging have not comprehensively reviewed the progress done in this novel research trend. Also, the significant increase in the number of works focused on the interpretation of the decision process of deep learning applied to medical imaging justifies the need for an updated review over the most recent methods not covered by the last surveys on the topic. {Moreover, the particular challenges of medical imaging analysis, including image complexity (anatomical structures, organs, and artifacts are often harder to identify compared to general images), data availability, and the misclassification risk, emphasize the need for a dedicated survey on interpretability applied to medical imaging.} 


To address these concerns, we comprehensively review the recent advances on explainable deep learning applied to medical diagnosis. In particular, this survey provides the following contributions: 

\begin{itemize}
    \item a review of the recent surveys on the topic of interpretable deep learning in medical imaging, including the major conclusions derived from each work, as well as a comparative analysis to our survey;
    \item an exhaustive list of the datasets commonly used in the study of interpretability of deep learning approaches applied to medical imaging;
    \item a comprehensive review of the state-of-the-art interpretable medical imaging approaches, covering both post-hoc and inherently interpretable models;
    \item a complete description of the metrics commonly used for benchmarking interpretability methods either for visual or textual explanations;
    \item a benchmark of interpretable medical imaging approaches regarding the quality of the textual explanations;
    \item the future research directions on the topic of interpretable deep learning in medical imaging.
\end{itemize}

\section{Related Surveys}
\label{sec:relatedword}

Explaining the decisions of deep learning models has been an active area of research, with various methods and algorithms proposed in the literature in the last years. The rapid pace of development of XAI has raised the need for comprehensive overviews of the advances in the state-of-the-art, and in most cases, the analysis of specific domains, due to the vast number of works published in the last years. Accordingly, in this section, we provide a critical analysis of the existing surveys in deep learning applied to medical imaging (section 2.1), with a particular focus on explainable approaches (section 2.2), and we compare the surveys analyzed with our survey (section 2.3).

\subsection{Deep Learning in Medical Image Analysis}

The advent of deep learning significantly changed the field of Computer Vision, where handcrafted feature extraction was replaced by end-to-end learning using Convolutional Neural Networks (CNNs). This new paradigm emerged in 2012 through the seminal work of Krizhevsky et al.~\cite{krizhevsky2012imagenet}, but it was not immediately incorporated by all the applications of Computer Vision. Litjens et al.~\cite{litjens2017survey} were the first to review the advances on medical image analysis fostered by the advent of deep learning, where is clear that the use of CNNs in medical imaging research has only become the standard approach in 2017, being clearly preferred over traditional handcrafted feature extraction for most of the anatomical regions. Based on the works reviewed, the authors concluded that data preprocessing and data augmentation techniques were essential to obtain superior results, and that the combination of medical images with text data (medical reports) could improve the image classification accuracy \cite{schlegl2015predicting}. 
Despite the relevance of this survey at the time, the rapid advances in the field of deep learning occurring in the last 5 years have made this work outdated, since the major conclusions of the survey are currently common sense, and novel deep learning models are currently being used in the medical imaging domain. 

Considering this, Rehman et al.~\cite{rehman2021survey} provided an updated overview of the advances on deep learning applied to medical image analysis. The survey was divided over different pattern recognition tasks (image classification, segmentation, image registration). Regarding the image classification task, the authors suggested using generative models to perform data augmentation to improve the results. The survey also gave future research directions to overcome the most common challenges identified by Litjens et al.~\cite{litjens2017survey}. The use of techniques such as transfer learning and synthetic data generation were suggested to address these challenges while improving the generalization capability of the developed strategies. Nevertheless, the authors concluded that the non-availability of large-scale annotated datasets remains one of the major challenges in medical imaging, which is impacting the performance of the deep learning models due to data overfitting and bias issues.

\subsection{Interpretable Deep Learning in Medical Imaging}

The works of Litjens et al.~\cite{litjens2017survey} and Rehman et al.~\cite{rehman2021survey} show that in the last years the use of deep learning has greatly improved the performance of medical imaging analysis algorithms, allowing also to create a myriad of approaches for the different image modalities and recognition tasks. Nevertheless, this contrasts with the adoption of these algorithms by clinicians who refuse to rely on decisions that they do not understand~\cite{lipton2017doctor}. In fact, as foresaw by Litjens et al.~\cite{litjens2017survey}, the importance of designing interpretable models for medical imaging has been growing in the last years and is nowadays one of the major challenges in medical imaging. The following paragraphs describe the different surveys focused on reviewing the recent advances on interpretable deep learning applied to medical imaging.  Additionally, the major conclusions derived from each work and a brief comparison to our survey are also outlined.

\subsubsection{General Reviews}

Tjoa and Guan~\cite{tjoa2020survey} provide a general overview of machine learning and deep learning interpretability methods with an emphasis on its application to medical field. The authors consider two types of interpretability: (i) perceptive interpretability, where the saliency methods are included, and (ii) interpretability by mathematical structures, which include mathematical formulations that can analyze the patterns in data. Although a significant number of works per image modality are covered, the survey lacks a comparison between the reviewed works. Also, the survey of Tjoa and Guan is more suitable for technical oriented readers, which is corroborated by a poor intuitive categorization for the medical community. Most works were discussed based on their mathematical foundation instead of describing the rationale behind the proposed method. As major conclusions, Tjoa and Guan state that combining visual and textual explanations is the most promising modality for conveying the explanations of medical imaging analysis algorithms, and that these algorithms should always be considered as a complementary aid/support to clinicians, who should be responsible for the final decision.

Singh et al.~\cite{singh2020explainable} propose a review of the works related to the explainability of deep learning models in the context of medical imaging. The methods are broadly divided in two major categories (attribution-based and non attribution-based) with respect to their capability of determining the contribution of an input feature to the target output. Both categories are reviewed by describing works applied to the different image modalities of medical data. Nevertheless, the survey focuses primarily on the attribution-based category, providing a superficial discussion of existing methods on the different categories of non-attribution methods, where inherently interpretable approaches are included. Based on the works reviewed, the authors conclude that leveraging patient record data and images can be an exciting research direction to push forward the performance of deep learning in medical imaging. When compared to our survey, the work of Singh et al. lacks a comprehensive analysis of inherently interpretable approaches, an analysis of the available medical imaging datasets, and the benchmarking  of the most prominent reviewed methods.

Recently, Salahuddin et al.~\cite{Salahuddin2022} review a set of interpretability methods which are grouped into nine different categories based on the type of explanations generated. They also discuss the problem of evaluating explanations and describe a set of evaluation strategies adopted to quantitatively and qualitatively measure the explanations' quality. Similarly to the other surveys, the authors also emphasize the importance of involving clinicians in designing and validating interpretability models to ensure the utility of the generated explanations. For the future perspectives, Salahuddin et al. claim that case-based and concept-learning models are promising interpretability models for being inherently interpretable and achieving similar performance to black-box CNNs. Despite being one of the most complete surveys on the topic, it lacks the description of most relevant datasets of the field as well as the benchmarking of most prominent approaches reviewed.

\subsubsection{Specific Image Modality Reviews}

In contrast to the above-mentioned works, the work of Pocevi{\v{c}}i{\={u}}t{\.{e}} et al. \cite{Poceviciute2020} is focused on a particular image modality. The XAI techniques devised for digital pathology are reviewed with respect to three criteria: 1) what is going to be explained (e.g., model predictions, predictions uncertainty); 2) explanation modality; 3) how the explanations are derived (e.g., perturbation-based strategies, interpretable network design). The authors point out the importance of developing a toolbox for objectively measuring the quality of explanations, as the lack of an evaluation framework remains an open problem in the XAI field. Additionally, the authors state that the use of counterfactual examples can enhance the interpretability of the methods.  When compared to our survey, this work has disregarded the textual explanation modality, focusing solely on visual explanations, either by visual examples or saliency maps.

Gulum et al.~\cite{gulum2021review} produce a review of the visual explainability techniques applied to cancer detection from Magnetic Resonance Imaging (MRI) scans. 
Contrary to the work of Pocevi{\v{c}}i{\={u}}t{\.{e}} et al. \cite{Poceviciute2020}, Gulum et al. discuss the strategies used for measuring the quality of explanations, but they only consider one metric to quantitatively evaluate the explanations. They also emphasize that there is a lack of studies that assess the explanation methods based on human evaluation. As future directions, Gulum et al. highlight the need for developing inherently interpretable approaches, as opposed to the traditional post-hoc strategy. Finally, the authors proposed the use of uncertainty estimation associated with model predictions to perceive how a model is confident in making a prediction. Despite its relevance, the work of Gulum et al. is specific to a particular image modality (MRI) and target disease (cancer). 

\subsubsection{Specific Explanation Modality Reviews}

While visual explanations are usually the primary option for explaining the model decisions, these strategies can be unreliable since they often highlight regions regardless of the class of interest~\cite{rudin2019stop}. This has fostered the research on textual explanations, and in the particular case of medical imaging led researchers to devise approaches capable of producing different types of textual explanations: 1) textual concepts and 2) textual reports. The recent developments on this topic have been covered in the surveys of Messina et al.~\cite{messina2022survey} and Ayesha et al. \cite{AYESHA2021107856}.

In~\cite{messina2022survey}, a thorough overview of the current state-of-the-art on automatic report generation from medical images is provided. The authors review 40 papers with respect to four dimensions: datasets used, model design, explainability and evaluation metrics. Furthermore, a benchmark of most relevant approaches is provided with respect to the performance in terms of NLP metrics on IU Chest X-ray dataset. Based on the works analysed, the authors identify the following challenges and future research directions: 1) the validation of the obtained explanations by clinicians is impractical, being necessary to create automatic metrics positively correlated with the clinicians opinion; 2) most research has concentrated on chest X-rays, mainly due to the availability of public data; 3) supervised learning may not be the most adequate strategy for medical report generation learning, whereas reinforcement learning seems a more reasonable training paradigm to explore.

Similarly to the work of Messina et al., Ayesha et al. \cite{AYESHA2021107856} presents a detailed survey of the existing automatic caption generation methods for medical images. The most used datasets and the evaluation metrics are also discussed. An extensive study is done around the most significant works under the various deep learning-based medical imaging caption generation methods, namely encoder-decoder based, retrieval-based, attention-based, concepts detection-based, and patient's metadata-based. Additionally, a comparative analysis of the performance of the methods reviewed is provided. Finally, Ayesha et al. suggest some future research directions to deal with the main open issues in medical imaging. They point out the lack of large-scale annotated datasets as the major limitation in the medical imaging field, where the data is scarce and often mislabelled. Also, they claim that the lack of a suitable evaluation metric to assess the generated caption remains an open problem since the evaluation of the generated text is still based on standard NLP metrics, such as BLEU score, ROUGE, METEOR, and CIDEr. As a final remark, Ayesha et al. also indicate the importance of having a model capable of detecting multiple diseases simultaneously.

\begin{table*}[t]
  \caption{Comparative analysis between the surveys on the topic of explainable deep learning applied to medical imaging. Our survey is the first to comprehensively review the advances on the topic regarding the different explanation modalities and the explanation processes. Also, it analyses the most relevant datasets on the the field, as well as their use for the development of explainable approaches.}
  \label{tab:surveys_comparision}
  \resizebox{0.9\textwidth}{!}{%
  \begin{tabular}{l|c|cc|cc|c|c}
    \toprule
    \multirow[t]{2}{*}{\textbf{Survey}} & \multirow[t]{2}{*}{\textbf{Year}} & \multicolumn{2}{c|}{\textbf{Explanations}} & \multicolumn{2}{c|}{\textbf{Model Type}} & \multirow[t]{2}{*}{\textbf{Medical Imaging }} & \multirow{2}{*}{\makecell{{\textbf{Benchmarking}} \\ {\textbf{Performance}}}} \\
    & & \textit{Visual} & \textit{Textual} &\textit{ Post-hoc }& \textit{In-model} & \textbf{Datasets} \\
    \midrule
    Pocevi{\v{c}}i{\={u}}t{\.{e}} et al. \cite{Poceviciute2020} & 2020 & \cmark & \xmark & \cmark & \cmark & \xmark & { \xmark }\\
    Tjoa and Guan \cite{tjoa2020survey} & 2020 & \cmark & \cmark & \cmark & \cmark & \xmark & { \xmark } \\
    Singh et al. \cite{singh2020explainable} & 2020 & \cmark & \cmark & \cmark & \cmark & \xmark &{ \xmark } \\
    Gulum et al. \cite{gulum2021review} & 2021 & \cmark & \xmark & \cmark  & \cmark & \xmark & { \xmark }\\
    Ayesha et al. \cite{AYESHA2021107856} & 2021 & \xmark & \cmark & \xmark & \cmark & \cmark & { \cmark }\\
    Salahuddin et al. \cite{Salahuddin2022} & 2022 & \cmark & \cmark & \cmark & \cmark & \xmark & { \xmark }\\
    Messina et al. \cite{messina2022survey} & 2022 & \xmark & \cmark & \xmark & \cmark & \cmark & { \cmark }\\
    \textbf{This survey} & {2023} & \cmark & \cmark & \cmark & \cmark & \cmark & { \cmark } \\
    
    \bottomrule
  \end{tabular}%
  }
\end{table*}

\subsection{Discussion}

Despite the significant contributions of each reviewed survey, few of them have described the most important datasets for medical imaging. Moreover, most surveys focused on particular aspects of interpretability, such as visual or textual approaches, and few works have comprehensively reviewed inherently interpretable models devised for medical image analysis. Another problem was the lack of a performance comparison among the reviewed methods. Accordingly, this survey covers these limitations by providing a broader overview of the current state-of-the-art XAI applied to medical diagnosis, including uni and multimodal approaches, followed by the most important medical imaging datasets and a comparative analysis of the models performance using standard evaluation metrics. {In addition, this survey explores a contemporary trend and an under-exploited category of inherently interpretable models, specifically concept-based learning approaches. As delineated in subsequent sections, these approaches are advantageous for medical diagnosis as they provide explanations in the context of high-level concepts that align with the knowledge of the physicians and promote the interaction between physicians and AI through model intervention.} Table \ref{tab:surveys_comparision} summarizes the major differences between the reviewed surveys and our work (This survey). 






\begin{figure}[t]
    \centering
    \includegraphics[width=\textwidth]{ 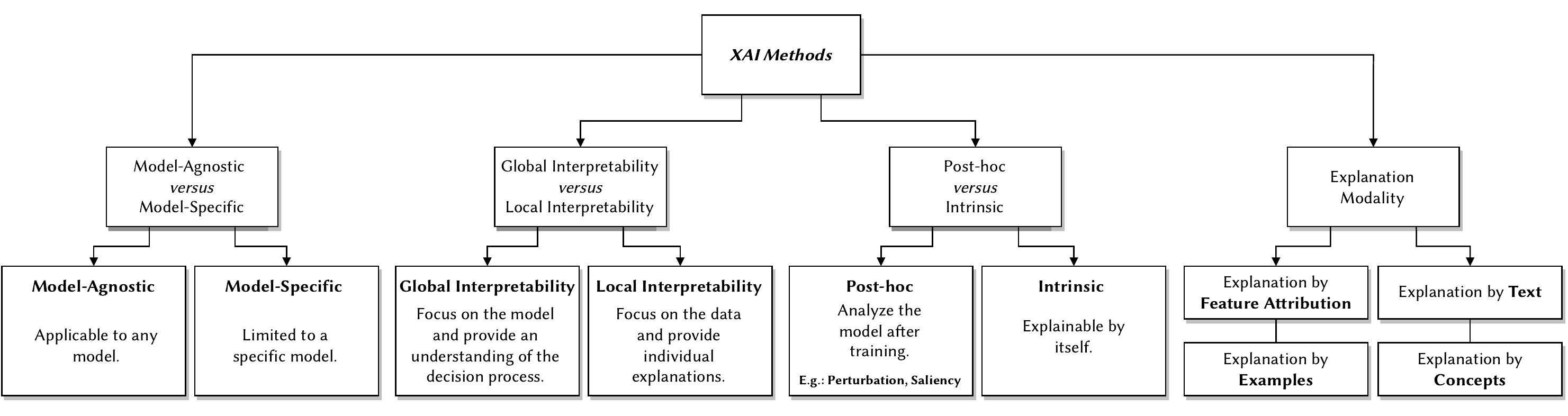}
    \caption{\textbf{Proposed categorization of the XAI methods taxonomy.} The proposed categorization was inspired by the various taxonomies presented in the reviewed papers.}
    \label{fig:xai_taxonomy}
\end{figure}

\section{Background in XAI}
\label{sec:xai_methods}

From a historical perspective, the problem of explaining expert systems has its origin in the mid-80s~\cite{moore1988explanation}, but the term XAI was only introduced in 2004 by Van et al.~\cite{van2004explainable}. Nevertheless, XAI has only gained prominence when deep learning dominated AI, and the first sign of this interest was demonstrated by the launching, in 2015, of the Explainable AI program by DARPA, whose primary goal was producing more explainable models to increase their understanding and transparency, leading to greater trust by users. Later, the European Union (EU)~\cite{goodman2017european} introduced legislation about the ``right to algorithmic explanation'' which provided citizens with the right to receive an explanation for algorithmic decisions obtained from personal data. Considering this, researchers shifted their efforts towards the creation of interpretable models rather than simply focusing on accuracy, leading to an exponential increase in the popularity and interest on XAI, whose number of works on the topic has rapidly increased in the last years.


This section provides a general overview of the taxonomy of XAI methods, the description of seminal XAI methods, as well as the existing frameworks providing implementations of these methods (Table \ref{tab:software} in appendix \ref{appx:int_frameworks}). 


\subsection{XAI Methods Taxonomy}

Based on the reviewed literature, XAI methods can be categorized according to three criteria: (i) Model-Agnostic versus Model-Specific; (ii) Global Interpretability versus Local Interpretability; and (iii) Post-hoc versus Intrinsic. Figure \ref{fig:xai_taxonomy} illustrates the general taxonomy of the XAI methods, and each category is detailed in the following paragraphs.

\paragraph{\textbf{Model-Agnostic versus Model-Specific}}

A distinguishing factor between interpretability approaches is their comprehensiveness regarding the models they can be applied to. Model-agnostic methods can be used to explain arbitrary models, not being limited to a specific model architecture. 
Conversely, model-specific methods are restricted to a specific model architecture, meaning that these methods require access to the model’s internal information. 

\paragraph{\textbf{Global Interpretability versus Local Interpretability}}

The type of explanations provided by XAI methods can be broadly divided into global and local whether the explanations provide insights about the model functioning for the general data distribution or for a specific data sample, respectively. Global interpretability methods explain which patterns in the data, i.e., class features, contributed the most to the model's prediction. These explanations can reveal critical reasoning about what the model is learning. On the other hand, local interpretability methods seek to explain why a model performs a specific prediction for a single input. 

\paragraph{\textbf{Post-hoc versus Intrinsic}}

This criterion distinguishes the methods with respect to whether the explanation mechanism lies in the internal architecture of the model (intrinsic) or if it is applied after the learning/development of the model (post-hoc). Post-hoc methods usually operate by perturbing parts of the data so that they can understand the contribution of different features in the model prediction, or by analytically determining the contribution of different features to the model prediction. On the other hand, intrinsic models, also known as in-model approaches or inherently interpretable models, are self-explainable since they are designed to produce human-understandable representations from the internal model features.

\paragraph{\textbf{Explanation Modality}}

{Explanation modality refers to the type of explanation provided by each interpretability method. Among the reviewed methods, the explanation can be provided in the form of saliency maps (Explanation by Feature Attribution), semantic descriptions (Explanation by Text), similar examples (Explanation by Examples), or using high-level concepts (Explanation by Concepts). In Chapter \ref{sec:methods}, we used this categorization to discuss the reviewed methods.}

\subsection{Classical XAI Methods}

The first attempts to explain deep learning models relied on the post-hoc analysis of the models. In spite of the criticism that post-hoc approaches have been recently subjected to~\cite{rudin2019stop}, they are still being used in many domains of medical imaging, and their understanding is important to explain the advances on the topic of interpretable deep learning. As such, the following sections briefly describe the most popular XAI algorithms according to the two major categories of post-hoc analysis.

\subsubsection{Perturbation-based methods}
\label{subseq:perturbation_based_methods}

The rationale behind perturbation-based methods is to perceive how a perturbation in the input affects the model's prediction. Examples of perturbation-based methods are LIME~\cite{ribeiro2016should} and SHAP~\cite{lundberg2017unified}. 

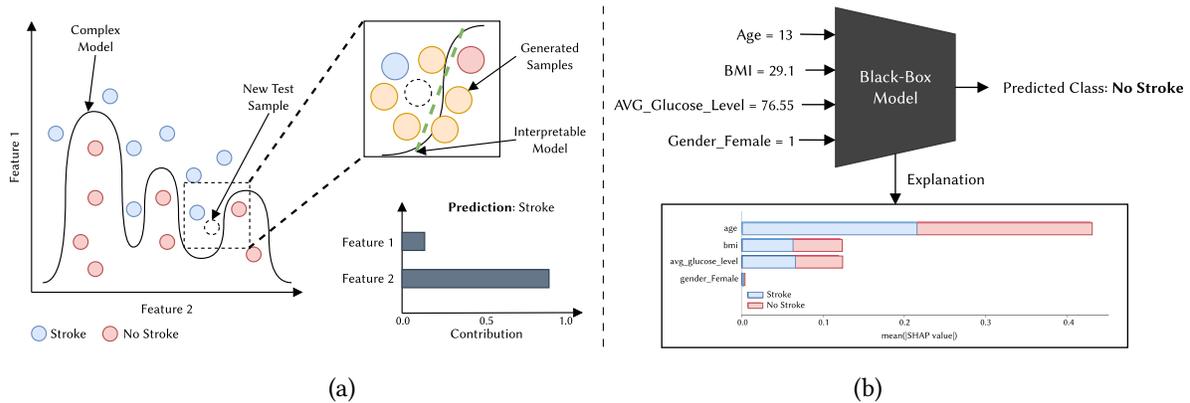
\begin{figure}[t]
\begin{tikzpicture}[font=\normalfont]

    \node (fig1) at (-0.15,0) 
    {\includegraphics[width=0.5\textwidth]{ 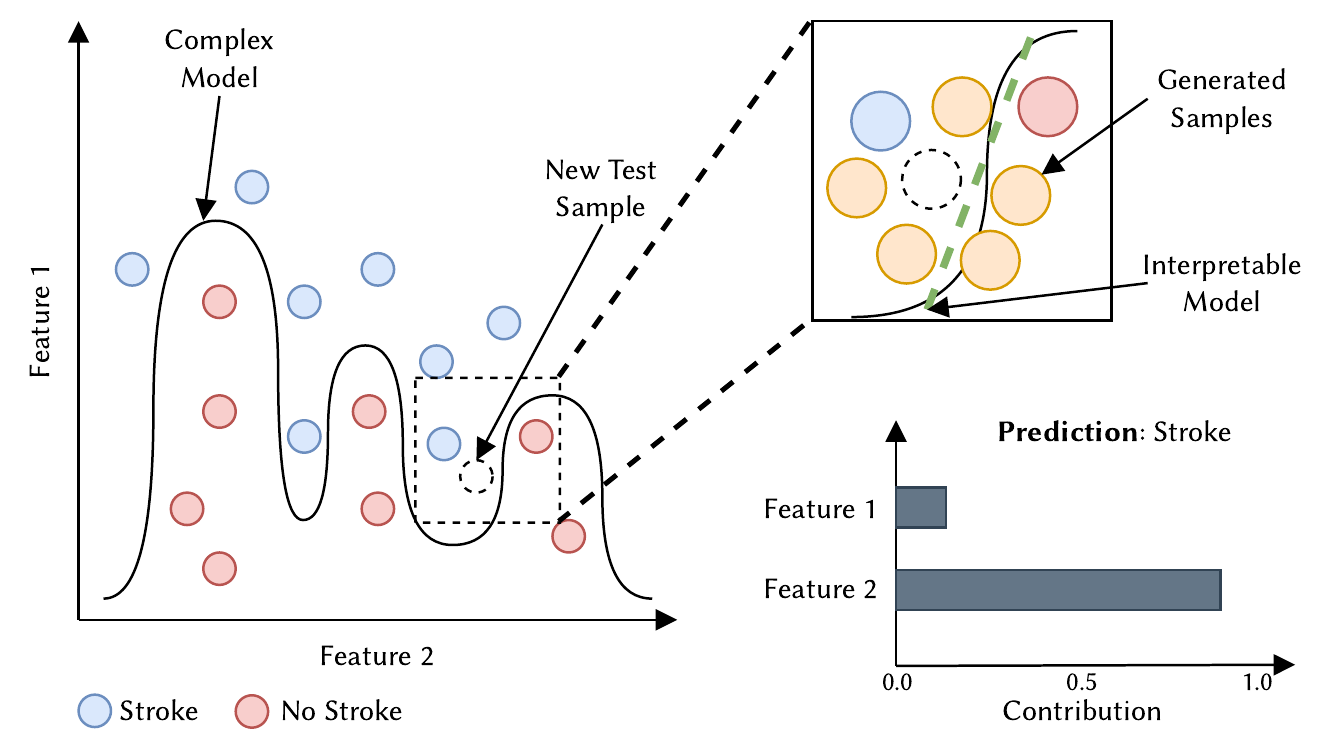}};
    \node (fig3) at (0.5\textwidth,0) 
    {\includegraphics[width=0.5\textwidth]{ 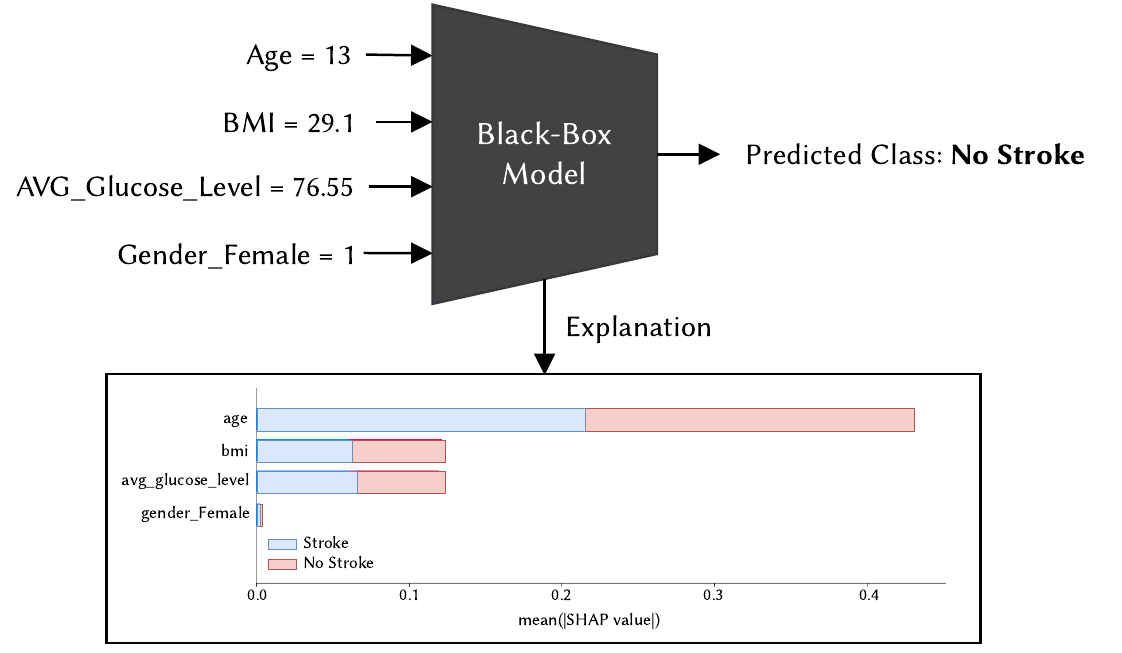}};
    
    \node at (0.03\textwidth,-2.8) {(a)};
    \node at (0.47\textwidth,-2.8) {(b)};
    
    \draw[dashed] (3.95,2.3) -- (3.95,-2.3);
    
\end{tikzpicture}
\caption{(a) \textbf{LIME}. The black curved line represents a decision boundary learned by the complex black-box model. LIME explains a new test sample (dashed circle), by fitting an interpretable model (represented by a green dashed line) to the variations of the test sample (orange circles), which are generated by randomly perturbing the test sample features. The fitted model allows to perceive the contribution of each feature for classifying that specific test sample. (b) \textbf{SHAP}. The predicted risk of stroke of a classification model for a 13 years-old female person, a body mass index of 29.1 and an average glucose level of 76.55 was ``No Stroke''. As evidenced by the bar plot, which provides the Shapley values for each feature, ``age'' was the feature with a higher impact on the prediction of ``No Stroke'', followed by the BMI and average glucose level features.}
\label{fig:lime_shap}
\end{figure}

\paragraph{\textbf{LIME}} LIME~\cite{ribeiro2016should} stands for Local Interpretable Model-agnostic Explanations. As the name suggests, it can explain any black-box model, and according to the XAI taxonomy is a post-hoc, model-agnostic method providing local explanations. The intuition behind LIME is to approximate the complex model (black-box model) locally with an interpretable model, usually denoted as local surrogate model. Thus, an individual instance is explained locally using a simple interpretable model around the prediction, such as linear models or decision trees. Figure \ref{fig:lime_shap}a) provides an intuitive illustration of the overall functioning of LIME.


In order to approximate the model prediction locally, a new dataset consisting of perturbed samples conditioned on their proximity to the instance being explained is used to fit the interpretable model. The labels for those perturbed samples are obtained through the complex model. In the case of tabular data, the perturbed instances are sampled around the instance being explained, by randomly changing the feature values in order to obtain samples both in the vicinity and far away from the instance being explained. Analogously, when LIME is applied to the image classification problem, the image being explained is first segmented into superpixels, which are groups of pixels in the image sharing common characteristics, such as colour and intensity. Then, the perturbed versions of the original data are obtained by randomly masking out a subset of superpixels, resulting in an image with occluded patches. The new dataset used to fit the interpretable model consists of perturbed versions of the image being explained, and the superpixels with the highest positive coefficients in the interpretable model suggest they largely contributed to the prediction. Thus, they will be selected as part of the interpretable representation that is simply a binary vector indicating the presence or absence of those superpixels.


\paragraph{\textbf{SHAP}} SHAP~\cite{lundberg2017unified} was inspired on the Shapley values from the cooperative game theory~\cite{ShapleyBook} and operates by determining the average contribution of a feature value to the model prediction using all combinations of the features powerset. As an example, given the task of predicting the risk of stroke based on age, gender and Body Mass Index (BMI), the SHAP explanations for a particular prediction are given in terms of the contribution of each feature. This contribution is determined from the change observed in model prediction when using the $2^n$ combinations from the features powerset, where the missing features are replaced by random values. Figure \ref{fig:lime_shap}b) illustrates the above-described example. Similarly to LIME, SHAP is a local model-agnostic interpretation method that can be applied to both tabular and image data. In the case of tabular data, the explanation is given in the form of importance values to each feature. In the case of image data, it follows a similar procedure to the LIME, by calculating the Shapley values for all possible combinations between superpixels. Several variations of SHAP method were proposed to approximate Shapley values in a more efficient way, namely KernelSHAP, DeepSHAP and TreeSHAP~\cite{lundberg2018consistent}.


\subsubsection{Saliency}

Saliency maps are one of the most popular techniques to explain the decisions of a model. Saliency methods produce visual explanation maps representing the importance of image pixels to the model classification. 

Class Activation Mapping (CAM)~\cite{zhou2016learning} is a seminal saliency method which allowed the generation of a saliency map using a linear combination of the output of the last Global Average Pooling (GAP) layer of the network. Despite being a seminal contribution, CAM can only be applied to architectures following a specific pattern. To address this problem, Selvaraju et al.~\cite{selvaraju2017grad} proposed the Gradient-weighted Class Activation Mapping (Grad-CAM)~\cite{selvaraju2017grad} that uses the gradient information of the target class with respect to the input image to produce a class-discriminative localization map that acts as a visual explanation for the model's prediction. Grad-CAM is, therefore, a generalization of CAM.
Alternatively, SmoothGrad~\cite{smilkov2017smoothgrad} is another gradient-based explanation method whose core idea is to attenuate the noise of the explanations provided by gradient-based techniques. The rationale behind SmoothGrad is to sample multiple images from the input image by adding noise to it. Then, the sensitivity maps are computed for each sampled image. The final map is the average of the sensitivity maps.

The Integrated Gradients (IG)~\cite{sundararajan2017axiomatic} is an attribution method that relies on generating a set of images between the baseline and the original image using linear interpolation. These interpolated images are minor changes in the feature space between the baseline and input image and consistently increase with each interpolated image’s intensity. Calculating the gradients per feature (pixels) makes it possible to measure the correlation between changes to a feature and changes in the model’s predictions. The pixels with a high score are the ones that contributed the most to the prediction. The Layer-wise Relevance Propagation (LRP)~\cite{bach2015pixel} is an alternative solution to the use of gradients, where the decision function is decomposed into the relevance score of each neuron in the network. The output is propagated backwards through the model to determine the relevance score of the input, allowing to produce an importance heatmap of image pixels.

The main advantage of saliency methods is the reduced computational cost when compared to perturbation-based methods. However, different studies argue that the explanations provided by gradient-based methods can be ambiguous and unreliable, as well as sensitive to adversarial perturbations~\cite{adebayo2018sanity, ghorbani2019interpretation}.

\section{Datasets}
\label{sec:datasets}

Among the reviewed literature, a set of 25 publicly available medical imaging datasets were considered {based on the reviewed papers} to provide a thorough overview of the existing medical imaging databases. Table \ref{tab:methods} presents the main characteristics of the selected datasets, grouped by image type.

\begin{table*}[h!]
  \caption{Medical Imaging Datasets. The datasets marked with a ``*'' have reports written in Spanish. The datasets marked with a ``**'' have reports written in Portuguese.}
  \label{tab:datasets}
  \resizebox{0.85\textwidth}{!}{%
  \begin{tabular}{lcccc}
    \toprule
    \textbf{Dataset} & \textbf{Image Type} & \textbf{Year} & \textbf{No. Images} & \textbf{Notes}\\
    \midrule
    IU Chest X-Ray~\cite{demner2016preparing}    & Chest X-ray & 2016 & 7,470 & Includes reports \\
    Chest X-Ray14~\cite{wang2017chestx} & Chest X-ray & 2017  & 112,120 & Multiple labels \\
    CheXpert~\cite{irvin2019chexpert}   & Chest X-ray & 2019 & 224,316 & Multiple labels \\
    MIMIC-CXR~\cite{johnson2019mimic}   & Chest X-ray  &  2019 & 377,110 & Includes reports \\
    PadChest$^{*}$~\cite{bustos2020padchest}  & Chest X-ray  & 2020 & 160,868 & Includes reports \\
    VinDr-CXR~\cite{nguyen2020vindrcxr} & Chest X-ray & 2020  & 18,000 & Multiple labels \\
    COVIDx~\cite{wang2020covid} & Chest X-ray & 2020 & 13,975 & Multiclass \\
    \midrule
    Inbreast$^{**}$~\cite{moreira2012inbreast} & Mammography X-ray &  2012 & 410 & Includes reports \\
    CBIS-DDSM~\cite{lee2017curated} & Mammography X-ray &  2017 & 2,620 & Multiclass \\
    \midrule
    VinDr-SpineXR~\cite{nguyen2021vindr}    & Spinal X-ray & 2021 & 10,469 & Multiple labels/BBox \\
    \midrule
    Knee Osteoarthirtis~\cite{nevitt2006osteoarthritis} & Knee X-ray & 2006 & 8,894 & Multiclass \\
    \midrule
    PH$^2$~\cite{mendoncca2013ph} & Dermatoscopic Images  & 2013 & 200 & Multiple labels/Lesion segment. \\
    HAM10000~\cite{tschandl2018ham10000} & Dermatoscopic Images & 2018  & 10,015 & Multiclass/Lesion segment. \\
    SKINL2~\cite{skinl2} & Dermatoscopic Images & 2019 & 376 & Multiclass \\
    Derm7pt~\cite{Kawahara_JBHI_2019} & Dermatoscopic Images & 2019 & 2,000 & Multiclass \\
    ISIC 2020~\cite{rotemberg2021patient} & Dermatoscopic Images  &  2020 & 33,126 & Multiple labels \\
    {SkinCon~\cite{daneshjou2022skincon}} & Dermatoscopic Images & 2022 & 3,230 & Concept annotations \\
    \midrule
    BreakHis~\cite{spanhol2015dataset}  & Microscopy Images & 2015 & 9,109 & Multiclass \\
    Camelyon17~\cite{litjens20181399}   & Microscopy Images & 2018 & 1,000 & Multiclass \\
    Databiox~\cite{bolhasani2020histopathological} & Microscopy Images  & 2020  & 922 & Multiclass \\
    BCIDR$^{(priv)}$~\cite{zhang2017mdnet} & Microscopy Images & 2017 & 5,000 & Includes reports \\
    \midrule 
    APTOS~\cite{aptos_dataset}  & Retina Images & 2019 & 5,590 & Multiclass \\
    \midrule
    LIDC-IDRI~\cite{armato2011lung} & CT scans  &  2011 & 1,018 & Includes annotations \\
    {COV-CTR~\cite{li2023auxiliary}} & CT scans & 2023 & 728 & Includes reports \\
    \midrule
    PEIR~\cite{jing2017automatic} & Photographs & 2017  & 33,648 &  Includes reports \\    ROCO~\cite{roco_dataset} & Multimodal & 2018 & 81,825 & Includes reports/UMLS Concepts\\
    \bottomrule
  \end{tabular}%
  }
\end{table*}

\subsection{Chest X-ray}

With respect to X-ray imaging modality, IU Chest X-ray~\cite{demner2016preparing}, Chest X-ray 14~\cite{wang2017chestx}, CheXpert~\cite{irvin2019chexpert}, MIMIC-CXR~\cite{johnson2019mimic}, PadChest~\cite{bustos2020padchest}, COVIDx~\cite{wang2020covid}, and VinDr-CXR~\cite{nguyen2020vindrcxr} datasets pertain to the chest anatomical region. Moreover, IU Chest X-Ray, MIMIC-CXR, and PadChest datasets include free-text radiology reports. The reports are written in English, except for the PadChest dataset, whose reports are written in Spanish. MIMIC-CXR and CheXpert are the largest databases, composed by $377,110$ and $224,316$ radiographs, respectively. CheXpert does not provide the raw free-text reports, but it provides an automated rule-based labeller for extracting keywords from medical reports conforming to the Fleischner Society’s recommended glossary~\cite{Fleischner}. This tool was also used in the MIMIC-CXR dataset to extract the labels from the radiology reports.
UI Chest X-Ray~\cite{demner2016preparing} comprises $7,470$ chest X-ray images jointly with $3,955$ free-text reports, being the most used dataset in the literature. The VineDr-CXR consists of $18,000$ chest X-ray images annotated with $22$ findings (local labels) and $6$ diagnosis (global labels). The local labels are inferred from the ``findings'' section in the radiology reports. In contrast, the global labels come from ``impressions'' section and indicate suspected diseases, such as ``Pneumonia'', ``Tuberculosis'', ``Lung Tumor'', ``Chronic obstructive pulmonary disease'', ``Other diseases'', and ``No Findings''. Additionally, each ``finding'' is annotated on the X-ray image with a bounding box. Finally, COVIDx~\cite{wang2020covid} dataset comprises 13,975 chest X-ray images across $13,870$ patient cases, distributed by ``Normal'', ``Pneumonia'' and ``COVID-19'' cases.

It is worth noting that the majority of the chest X-ray dataset's labels were extracted using an automatic rule-based labeler, such as the CheXpert NLP tool~\cite{irvin2019chexpert}. However, relying on these automated tools can pose several issues concerning the quality of the labels. Consequently, the authors of the VinDr-CXR~\cite{nguyen2020vindrcxr} dataset provided only radiologist-level annotations in both training and test sets. 

\subsection{Other X-ray modalities}

In the same segment of the X-ray imaging, and similarly to the VinDr-CXR~\cite{nguyen2020vindrcxr} dataset approach, VinDr-SpinalXR~\cite{nguyen2021vindr} is a recent dataset comprising $10,469$ spine X-ray images manually annotated by an experienced radiologist with bounding-boxes around abnormal findings in $13$ categories. The Knee Osteoarthritis~\cite{nevitt2006osteoarthritis} dataset contains 8,894 knee X-rays for both knee joint detection and knee Kellgren and Lawrence grading~\cite{klgrading}, whose value ranges from 0 to 4, according to the level of severity. Regarding the mammography datasets, Inbreast~\cite{moreira2012inbreast} consists of $410$ mammography X-rays along with 115 radiology reports written in Portuguese. Similarly, CBIS-DDSM~\cite{lee2017curated} dataset is composed of $2,620$ mammography scans distributed into ``Normal'', ``Benign'', and ``Malignant'' cases. 

\subsection{Dermatoscopy}

In the scope of dermatology, the ISIC 2020~\cite{rotemberg2021patient} dataset consists of $33,126$ skin lesion images of different categorizations (malignant, melanoma, keratosis, etc), and it is part of the International Skin Imaging Collaboration, which promotes annual challenges to enhance the diagnosis of malignant skin lesions in dermoscopy images. The HAM10000~\cite{tschandl2018ham10000} dataset consists of $10,015$ dermatoscopic images distributed into diagnostic categories in the realm of pigmented lesions. The PH$^2$~\cite{mendoncca2013ph} dataset comprises $200$ dermoscopic images of melanocytic lesions, including common nevi, atypical nevi, and melanoma. Moreover, the PH$^2$ database includes medical segmentation of the lesions, clinical and histological diagnosis, and the assessment of several dermoscopic criteria. The Derm7pt~\cite{Kawahara_JBHI_2019} is another dermoscopic image dataset with over $2,000$ images annotated following the 7-point melanoma checklist criteria. Finally, the SKINL2~\cite{skinl2} database consists of a total of $376$ light fields of skin lesions manually annotated under eight categories, based on the type of skin lesion and using the correspondent International Classification of Diseases (ICD) code. {SkinCon~\cite{daneshjou2022skincon} includes 3,230 images from the Fitzpatrick 17k skin disease dataset~\cite{groh2021evaluating} annotated with 48 clinical concepts.}

\subsection{Microscopy}

With regard to datasets composed of microscopy images, the BreakHis~\cite{spanhol2015dataset} dataset comprises $9,109$ microscopic images of breast tumour tissue distributed in various magnifying factors (40x, 100x, 200x, and 400x) and categorized into ``benign'' and ``malignant'' tumours. The Camelyon 17~\cite{litjens20181399} dataset consists of $1,000$ annotated whole-slide images (WSI) of lymph nodes. It is part of a challenge whose primary goal is the classification of breast cancer metastases in WSI of histological lymph node sections. Similarly, Databiox ~\cite{bolhasani2020histopathological} dataset comprises $922$ histopathological microscopy images with four levels of magnification (4x, 10x, 20x, and 40x) for the task of Invasive Ductal Carcinoma (IDC) grading using three grades of IDC.

\subsection{Others}

For blindness detection purposes, the APTOS~\cite{aptos_dataset} dataset provides $5,590$ images of the retina taken using fundus photography. A clinician annotated each image according to the severity of diabetic retinopathy on a scale of 0 (Diabetic Retinopathy) to 4 (Proliferative Diabetic Retinopathy). {In the scope of COVID-19, COVID-19 CT Reports (COV-CTR)~\cite{li2023auxiliary} is a dataset composed of 728 CT scans paired with Chinese and English reports.}

Regarding databases with multiple imaging modalities, the Pathology Education Informational Resource (PEIR) is a multidisciplinary public access image database intended for medical education. The PEIR database consists of $33,648$ images and the respective descriptions from different sub-classes of PEIR albums (PEIR Pathology, PEIR Radiology, and PEIR Slice). Similarly, the Radiology Objects in COntext (ROCO)~\cite{roco_dataset} dataset is a multimodal image dataset consisting of $81,825$ radiology images divided into CT, MRI, X-ray Ultrasound, and Mammography. Each image is accompanied by the corresponding caption and the annotated Unified Medical Language System (UMLS) concepts.

\subsection{Discussion} 
\label{subsec:discussion_datasets}

According to Table \ref{tab:datasets}, Chest X-ray is the most popular imaging modality regarding simultaneously the number of datasets and their scale. {Large-scale datasets have the advantage that they can be used to train a model from scratch. However, the automatic labeling process of some examples could compromise the reliability of the model since some class labels are not human verified and can be mislabeled.} On the other hand, as evidenced by the number of images per dataset, the scarcity of large-scale datasets concerning other imaging modalities is noticed, {hindering the emergence of specialized task-specific models due to the limited training data.} {The majority of the medical imaging datasets have been gathered primarily for use in classification or segmentation tasks, where the labeling of class and mask annotations are sufficient for their intended purpose. Nevertheless, when these sets are utilized for interpretability purposes, their annotations may prove inadequate for a comprehensive qualitative evaluation. Nonetheless, some datasets, such as SkinCon and others, have been created explicitly with interpretability in mind and contain appropriate annotations to facilitate such evaluations. Accordingly, it is particularly important that researchers consider interpretability issues when building novel medical datasets, ensuring that appropriate annotations are included to further enable comprehensive and informative analyses of the data.}

\section{XAI Methods in Medical Diagnosis}
\label{sec:methods}

As aforementioned, deep learning models must confer transparency and trustworthiness when deployed in real-world scenarios. Moreover, this requirement becomes particularly relevant in clinical practice, where a less informed decision can put patient's lives at risk. Among the reviewed literature, several methods have been proposed to confer interpretability in the deep learning methods applied to medical diagnosis. The following sections summarize and categorize the most relevant works in the scope of explainable models applied to medical diagnosis. Furthermore, we give particular attention to the inherently interpretable neural networks and their applicability to medical imaging. We categorize the methods according to the explanation modality: (i) Explanation by Feature Attribution, (ii) Explanation by Text, (iii) Explanation by Examples, (iv) Explanation by Concepts, and (v) Other Approaches; inspired by the taxonomy presented in~\cite{molnar2022}. 
The list of the reviewed methods categorized by the interpretability method employed, image modality, and the dataset is provided in Table \ref{tab:methods_alltogether} in appendix \ref{appx:methods}\footnote{An interactive version of the Table 5 is provided in this \href{https://github.com/CristianoPatricio/Explainable-Deep-Learning-Methods-in-Medical-Image-Classification-A-Survey}{link}.}. 


\begin{figure}[t]
\begin{tikzpicture}[font=\normalfont]

    \node (fig1) at (-0.15,0) 
    {\includegraphics[width=0.5\textwidth]{ 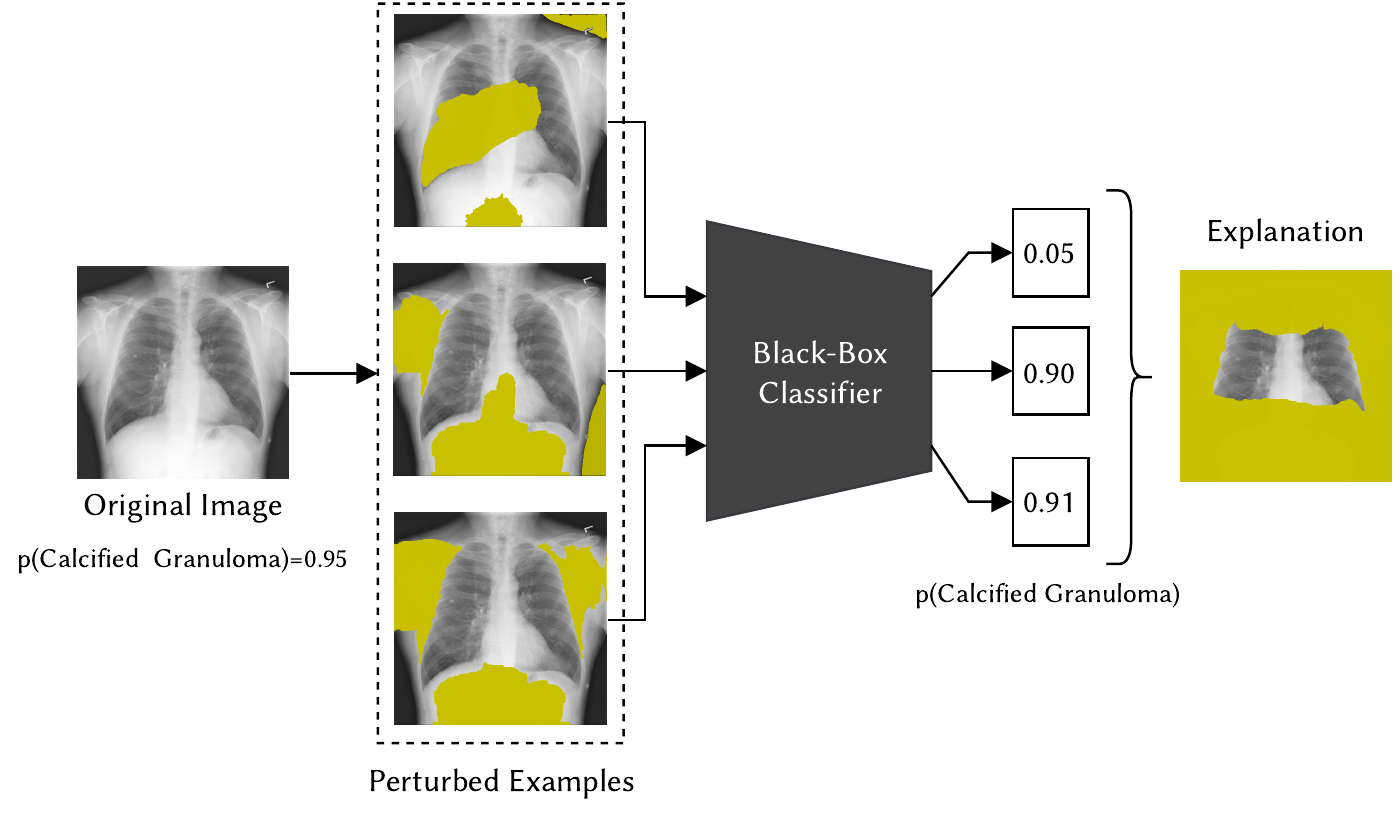}};
    \node (fig3) at (0.5\textwidth,0) 
    {\includegraphics[width=0.49\textwidth]{ 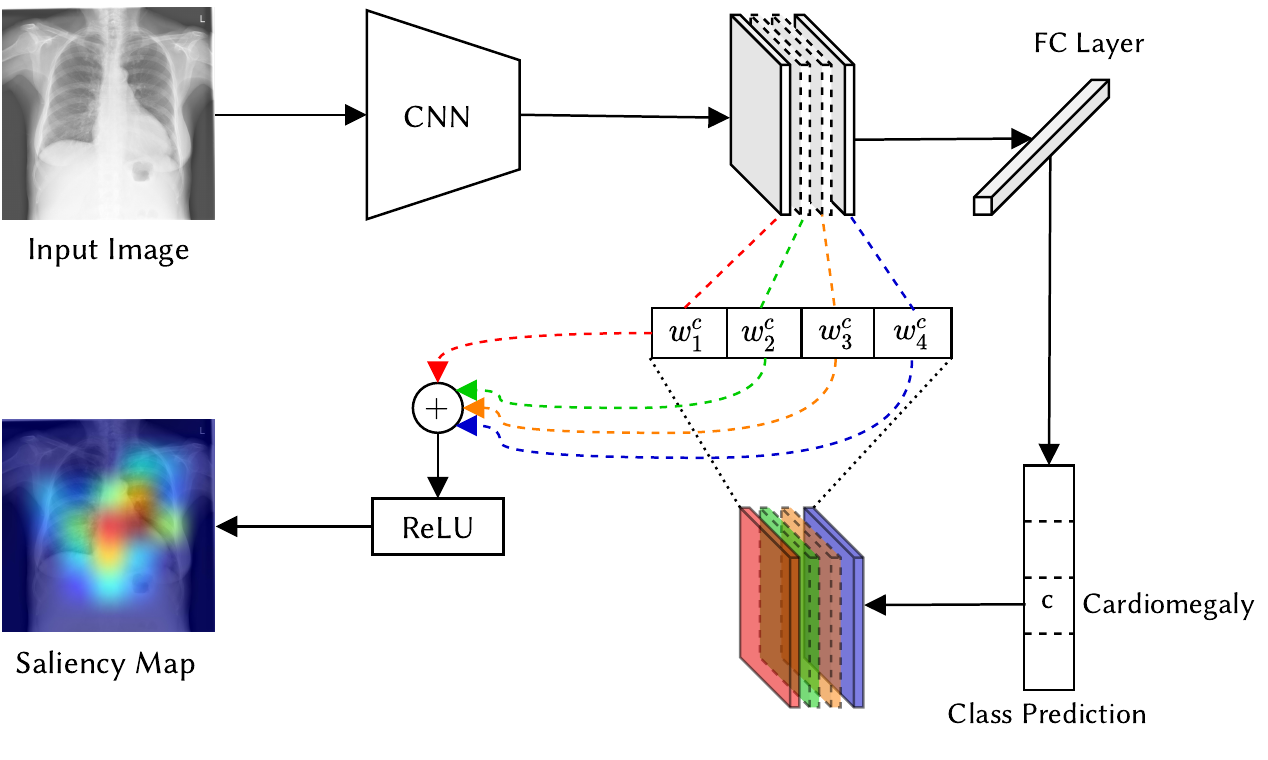}};
    
    \node at (0.03\textwidth,-2.25) {(a)};
    \node at (0.47\textwidth,-2.25) {(b)};
    
    \draw[dashed] (3.95,2.3) -- (3.95,-2.3);
    
\end{tikzpicture}
\caption{(a) \textbf{Perturbation-based methods}. The input image is randomly perturbed by turning on and off certain pixels, resulting in an image with occluded parts (\textit{Perturbed Examples} in the figure). Then, the perturbed image is fed to the classification model and the prediction confidence is exploited to determine the regions that contributed positively to the class prediction. Those regions will be considered to obtain the final explanation map (\textit{Explanation} in the figure). (b) \textbf{Saliency methods}. The input image is fed to the classification model to obtain a class prediction. Then, the gradient is calculated for the score of the class concerning the feature maps of the last convolutional layer. After calculating the importance of the feature map regarding the predicted class, they are weighted with each of respective weight, followed by a ReLU operation to obtain the final saliency map.}
\label{fig:feature_attribution}
\end{figure}
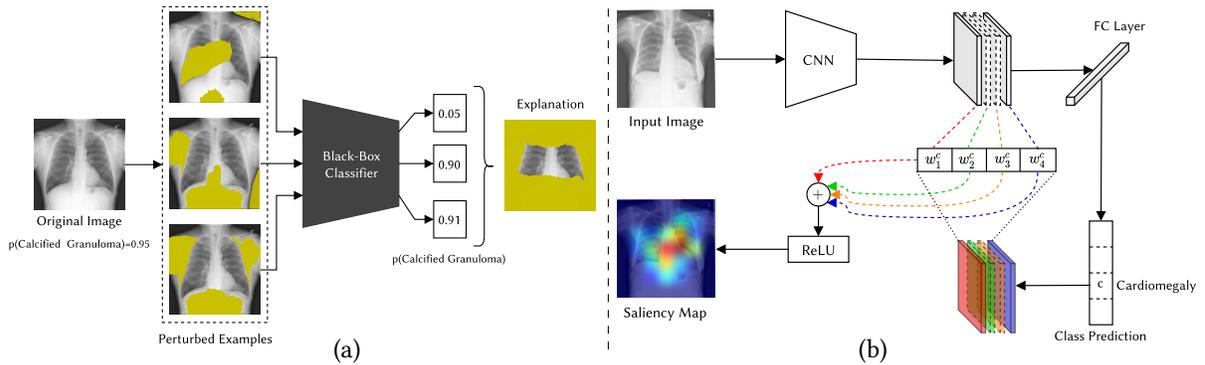

\subsection{Explanation by Feature Attribution}

Feature attribution methods indicate how much each input feature contributed to the final model prediction. These methods can work on tabular data or image data by depicting feature importance scores in a bar chart or using a saliency map, respectively. Among the existing feature attribution approaches in the literature, we categorize them into (i) perturbation-based methods and (ii) saliency methods, emphasizing their application to medical image analysis. A schematic diagram illustrating the general pipeline of these methods is shown in Figure \ref{fig:feature_attribution}.

\subsubsection{Perturbation-based methods}

As previously described in section \ref{subseq:perturbation_based_methods}, perturbation-based methods aim to perform a modification in the input data to perceive how it affects the model's prediction. Popular examples of perturbation-based methods are LIME~\cite{ribeiro2016should} and SHAP~\cite{lundberg2017unified}. Regarding the use of perturbation methods to explain the prediction of medical diagnosis algorithms, Malhi et al.~\cite{malhi2019explaining} applied the LIME method to explain the decisions of a classifier to detect bleeding in gastral endoscopy images. Similarly, Punn et al.~\cite{punn2021automated} applied the LIME technique to explain the predictions of various state-of-the-art deep learning models used to classify pulmonary diseases in chest X-ray images. Magesh et al.~\cite{magesh2020explainable} also used LIME to justify the decisions of a Parkison's disease classifier model. In the context of melanoma detection, Young et al.~\cite{Young_2019} used Kernel SHAP~\cite{lundberg2017unified} interpretability method to investigate its reliability in providing explanations for a melanoma classifier from dermoscopy images. They concluded that the interpretability strategy highlighted irrelevant features to the final prediction. The authors also conducted sanity checks on the interpretability methods, which confirmed that these methods often produce different explanations for models with similar performance. This can be explained by the fact that the model can learn some spurious correlations, causing interpretability methods to give exaggerated importance to those spurious regions highlighted on the produced saliency maps. In the same context, Wang et al.~\cite{wang2021interpretability} proposed a multimodal CNN for skin lesion diagnosis, which considered patient metadata and the skin lesion images. To analyze the contribution of each feature regarding the patient metadata, they adopted SHAP. Similarly, Eitel et al.~\cite{eitel2019testing} relied on an occlusion-based interpretability method~\cite{zeiler2014visualizing} and investigated its robustness for the task of Alzheimer’s disease classification. 

RISE was proposed by Petsiuk et al.~\cite{petsiuk2018rise}, consisting also of a post-hoc model-agnostic method for explaining the predictions of a black-box model through the generation of a saliency map indicating the important pixels to the model's prediction. The core idea is to probe the model with a set of perturbed images from the input image via random masking to perceive the model's response as important regions of the image are randomly occluded. The final saliency map is generated as a linear combination of the generated masks weighted with the output probabilities predicted by the model. For evaluation, the authors proposed two novel metrics, namely deletion and insertion, based on the removal and insertion of important pixels in the image to perceive the increase or the decrease of the model's performance.

\subsubsection{Saliency Methods}

Saliency methods allow to produce a saliency map in which each pixel is assigned a value that represents its relevance to the prediction of a certain class. Popular techniques are CAM~\cite{zhou2016learning}, Grad-CAM~\cite{selvaraju2017grad}, DeepLIFT~\cite{pmlr-v70-shrikumar17a} and Integrated Gradients~\cite{sundararajan2017axiomatic}.

Rajpurkar et al.~\cite{rajpurkar2018deep} proposed the CheXNeXt model to detect pulmonary pathologies and used CAM to identify the locations on the chest radiograph that contributed most to the final model prediction.

In the context of detecting COVID-19 from chest radiographs, Lin et al.~\cite{lin2020covid} used Grad-CAM and Guided Grad-CAM to investigate the regions that the model considered more discriminative. The produced heatmaps showed that when no preprocessing is used, the CNN tends to concentrate on non-lung areas (e.g., spine, heart, background), deemed irrelevant for the classification decision. However, when a masking process is used to highlight only the lung's area, the produced heatmaps highlight only the relevant regions since the CNN attention is limited to the critical area for detecting pulmonary diseases (lung's area). Following the same procedures, Lopatina et al.~\cite{lopatina2020investigation} used DeepLIFT attribution algorithm to investigate the decisions of a multiple sclerosis classification model, and Sayres et al.~\cite{sayres2019using} used Integrated Gradients to provide explanations for the task of predicting diabetic retinopathy from retinal fundus images.

In contrast to the previous approaches, Rio-Torto et al.~\cite{rio2020understanding} proposed an in-model joint architecture composed of an explainer and a classifier to produce visual explanations for the predicted class labels. The explainer consists of an encoder-decoder network based on U-Net, and the classifier is based on VGG-16. Since the classifier is trained using the explanations provided by the explainer, the classifier focuses only on the relevant regions of the image containing the class. The qualitative assessment of the provided explanations was carried out by using state-of-the-art explainability methods provided by the Captum library~\cite{captum}. Additionally, a quantitative analysis was also provided in terms of accuracy, average precision, AUROC, AOPC~\cite{samek2016evaluating} and the proposed POMPOM metric.

\subsubsection{Discussion} 

Despite the simplicity of the feature attribution methods and their applicability to a wide range of approaches, these methods may often produce ambiguous explanations, difficulting their qualitative evaluation. Furthermore, preprocessing techniques were required for some methods to generate more plausible explanations~\cite{lin2020covid}. Thus, researchers began exploring other modalities, such as textual explanations, and it was discovered that textual explanations were indeed valid explanations and, in some cases, preferred over visual explanations since they are inherently understandable by humans ~\cite{vandervelden2021explainable}. 


\subsection{Explanation by Text}

\begin{figure}[t]
    \centering
    \includegraphics[width=0.85\textwidth]{ 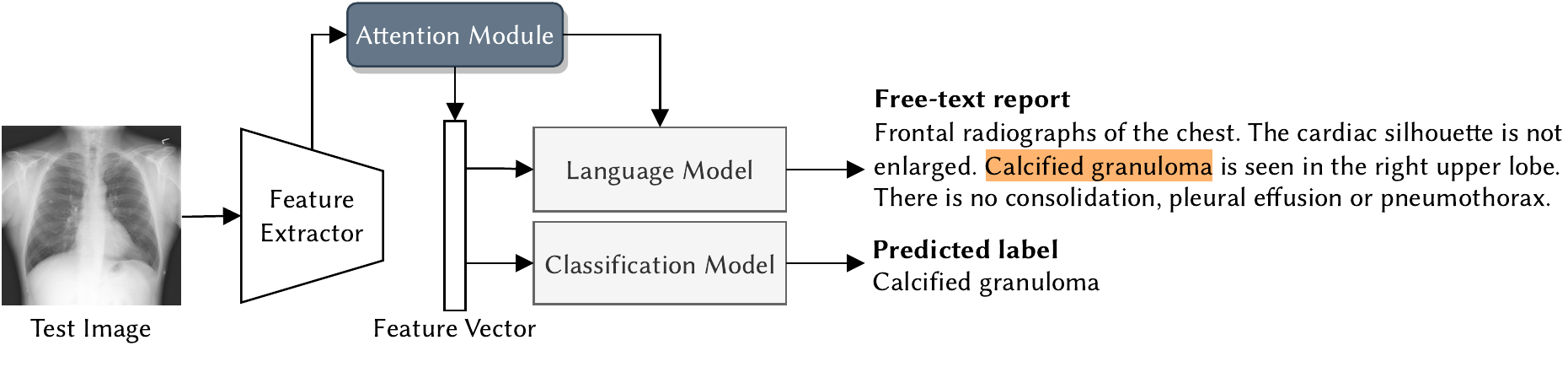}
    \caption{\textbf{Explanation by textual descriptions}. The typical architecture for obtaining a textual description from image data combines an image embedding model (e.g., CNN) for extracting the features from the input image and a language model (e.g., LSTM) for generating the word sentences. The attention module can be inserted between those two models to guide the language model to focus only on relevant regions of the input image to improve the generation of the word sentences.}
    \label{fig:explanation_by_text}
\end{figure}

The use of semantic descriptions became another way of explaining the model decisions since most of the clinicians prefer textual explanations compared to visual explanations solely, and the combination of textual and visual explanations over either alone ~\cite{gale2018producing}. In general, providing textual explanations can be built on three paradigms: (i) image captioning, (ii) image captioning with visual explanation, and (iii) concept attribution~\cite{vandervelden2021explainable}. Figure \ref{fig:explanation_by_text} depicts the general scheme adopted by most works to generate a textual description based on the visual features of the input image.

\subsubsection{{Image Captioning}}
The task of generating a textual description for explaining a model decision can be viewed as an {extension of the} image captioning problem, commonly treated in Natural Language Processing (NLP). Indeed, the vast majority of works that aim to generate a textual description for a given input image follow the classical strategy of combining a CNN for extracting the visual features with an RNN (e.g., LSTM) to generate the word sentences. Based on this paradigm, Sun et al.~\cite{Sun2019} developed a joint framework to generate sequences of words to provide a textual explanation for the task of diagnosing malignant tumors from breast mammography. Similarly, Singh et al.~\cite{singh2019chest} built on an encoder-decoder framework composed of a CNN and a stacked Long Short-Term Memory (LSTM) for automatically generating radiology reports from chest X-rays. Regarding image captioning with visual explanation, Zhang et al.~\cite{zhang2017mdnet} proposed a multimodal approach, dubbed MDNet, composed of an image embedding model and a language model, that can generate diagnostic reports, retrieve images by symptom descriptions, and visualize network attention. The MDNet model was evaluated on a dataset (BCIDR) containing histopathological images of bladder cancer. Furthermore, MDNet inspired several approaches. For example, Jing et al.~\cite{jing2017automatic} proposed a multi-task learning framework with a co-attention mechanism to guide the generation of text according to the localized regions containing abnormalities. They showed that a hierarchical LSTM model performs better in generating long text reports. Subsequently, Wang et al.~\cite{wang2018tienet} proposed TieNet, which makes use of attention modules to extract the most important information from chest X-ray images and also use their diagnostic reports in order to guide the model to produce more coherent reports. Similarly, Barata et al.~\cite{barata2019deep} proposed a hierarchical classification model that uses attention modules, including channel and spatial attention, to identify relevant regions in the skin lesions and subsequently guide further the LSTM attending at different locations whilst conferring more transparency to the network. Lee et al.~\cite{lee2019generation} also explained the decisions of a breast masses classifier using both visual and textual explanations, based on a CNN-RNN architecture. In the same fashion, Gale et al.~\cite{gale2018producing} proposed a model-agnostic interpretable method based on an RNN to produce textual explanations for the decisions of deep learning classifiers. Furthermore, they developed a visual attention mechanism in charge of highlighting the relevant regions for classifying hip fractures in pelvic X-rays. Yin et al.~\cite{yin2019automatic} also used attention mechanisms to attend to the regions at sentence level. They proposed the Hierarchical Recurrent Neural Network (HRNN) model, composed of two-level LSTMs: a word RNN and a sentence RNN. The sentence RNN produces the topic vectors whereas the word RNN receives the output of the sentence RNN and infers the words that constitute the final report. Moreover, they introduced a matching mechanism to map the topic vectors and the sentences into a jointly semantic space that minimizes a contrastive loss. Similar to the work of Yin et al.~\cite{yin2019automatic}, the model proposed by Liu et al.~\cite{liu2019clinically} generates topics from images and then completes sentences from these topics. When compared to ~\cite{yin2019automatic}, this work allows the generation of more coherent report generation due to the use of a fine-tuning process that uses reinforcement learning via CIDEr.

A more disruptive approach was introduced by Li et al.~\cite{li2018hybrid} that proposed the Hybrid Retrieval-Generation Reinforced Agent (HRGR-Agent) consisting of a retrieval policy module and a generation module. The retrieval policy module is responsible for deciding whether sentences are obtained from a generation module or retrieved from the template database, which is composed of a set of template sentences. Moreover, the retrieval policy and generation modules are updated via reinforcement learning, guided by sentence-level and word-level rewards using the CIDEr.

In contrast to the previous approaches, Chen et al.~\cite{chen2020generating} exploited the Transformer~\cite{vaswani2017attention} architecture, where they incorporated two memory modules into the decoder. These modules are responsible for memorizing textual patterns and assisting the decoder of the Transformer in generating radiology reports containing relevant information associated with chest X-ray images. {The same tendency is reflected in recent works~\cite{liu2021exploring,liu2021ACL,wang2022medical,wang2023metransformer, yang2023radiology}, which employ transformer-based models with additional custom modules to better capture the relevant features of input images, leading to an improved performance in the task of radiological report generation. Recently, Selivanov et al.~\cite{selivanov2023medical} introduced a novel image captioning architecture that combines two language models, incorporating image-attention (SAT)~\cite{xu2015show} and text-attention (GPT-3)~\cite{brown2020language}, resulting in an outstanding performance compared to the previous methods. For a more comprehensive analysis of the use of Transformers in medical imaging, we refer the reader to the survey of Shamshad et al.~\cite{shamshad2023transformers}.}

\subsubsection{Concept Attribution}

\begin{figure}[t]
    \centering
    \includegraphics[width=0.79\textwidth]{ 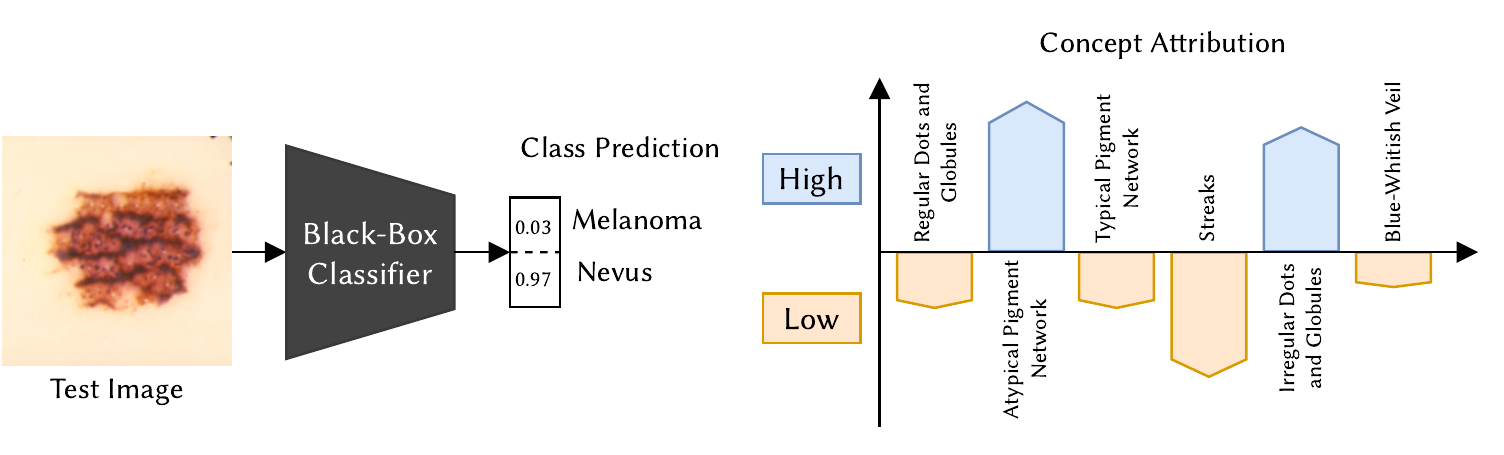}
    \caption{\textbf{Explanation by concept attribution}. In the first phase, the human-defined concepts (Regular Dots and Globules, Atypical Pigment Network, Typical Pigment Network, Streaks, Irregular Dots and Globules, and Blue-Whitish Veil) are modeled as numeric features, following the CAV technique. Then, after the input image passes through a classifier model, the class prediction is globally explained based on the importance of each concept to the final prediction.}
    \label{fig:concept_attribution}
\end{figure}

The idea behind concept-based attribution is learning human-defined concepts from the internal activations of a CNN. The use of concepts to provide global explanations was proposed by Kim et al. ~\cite{kim2018interpretability} with the introduction of the Concept Activation Vectors (CAVs), which provide explanations in terms of human-understandable concepts that are typically related to parts of the image. Kim et al. also proposed Testing with CAVs (TCAV), which enable to quantify the importance of a user-defined concept to a classification result. Figure \ref{fig:concept_attribution} illustrates the typical pipeline of concept-based attribution methods. In the context of medical imaging, the term ``microaneurysm'' can be viewed as a concept, being possible to be identified by humans in fundus imaging, and which denote the presence of diabetic retinopathy ~\cite{vandervelden2021explainable}. In the same line of research, Graziani et al.~\cite{graziani2020concept} proposed a framework for concept-based attribution to generate explanations for CNN decisions of a breast histopathology classifier. They built on TCAV by incorporating Regression Concept Vectors (RCV), which provide continuous-values measures of a concept instead of solely indicating its presence or absence. This is particularly useful in the medical domain since a value indicating, for example, tumor size is more informative than a binary value indicating its presence or absence. Graziani et al. also concluded that the learning of concepts by an intermediate layer of a CNN could be improved by removing spatial dependencies of the convolutional layers and introducing L2 norm regularization in the regression problem. Recently, Lucieri et al.~\cite{lucieri2022exaid} introduced ExAID, a framework that provides multimodal concept-based explanations for the task of melanoma classification. Their framework relied on CAVs for concept identification and used the TCAV method to estimate the influence of a specific concept on the decision. 
The authors provided textual explanations in the form of template phrases by only replacing the identified concepts and their importance to the prediction in the phrase structure. In order to localize the regions of the learned concepts in the latent space of the trained classifier, the authors used the Concept Localization Maps~\cite{lucieri2020explaining} which uses perturbation-based concept localization to generate a saliency map highlighting the relevant regions with respect to the learned concepts, thus providing visual explanations.

\subsubsection{Discussion}

In summary, except for the work of Chen et al.~\cite{chen2020generating}, all the explanation text-based approaches mentioned above rely on RNN architectures to generate text descriptions towards providing a more human-interpretable explanation for a model decision. However, as stated by Pascanu et al.~\cite{pascanu2013difficulty}, RNN-based approaches, such as LSTM, have some limitations in generating long text reports. On the contrary, concept-based attribution methods provide a more objective and human-understandable way of interpreting classification decisions. However, the main limitation of these methods is the need for manual annotations of the concept examples, which may be impractical for specific medical image modalities, increasing the need for involving clinicians in the annotation tasks. 



\subsection{Explanation by Examples}

The type of methods that explain a model decision by selecting a set of similar examples are dubbed example-based explanation methods. Apart from explaining algorithm decisions, this strategy is also commonly used between clinicians to explain the rationale behind their decision process. Below, we categorize example-based explanation methods into the following categories: (i) Case-Based Reasoning, (ii) Counterfactual Explanations, and (iii) Propotypes. 

\subsubsection{Case-Based Reasoning}

\begin{figure}[!h]
\begin{tikzpicture}[font=\normalfont]

    \node (fig1) at (-0.2,0) 
    {\includegraphics[width=0.5\textwidth]{ 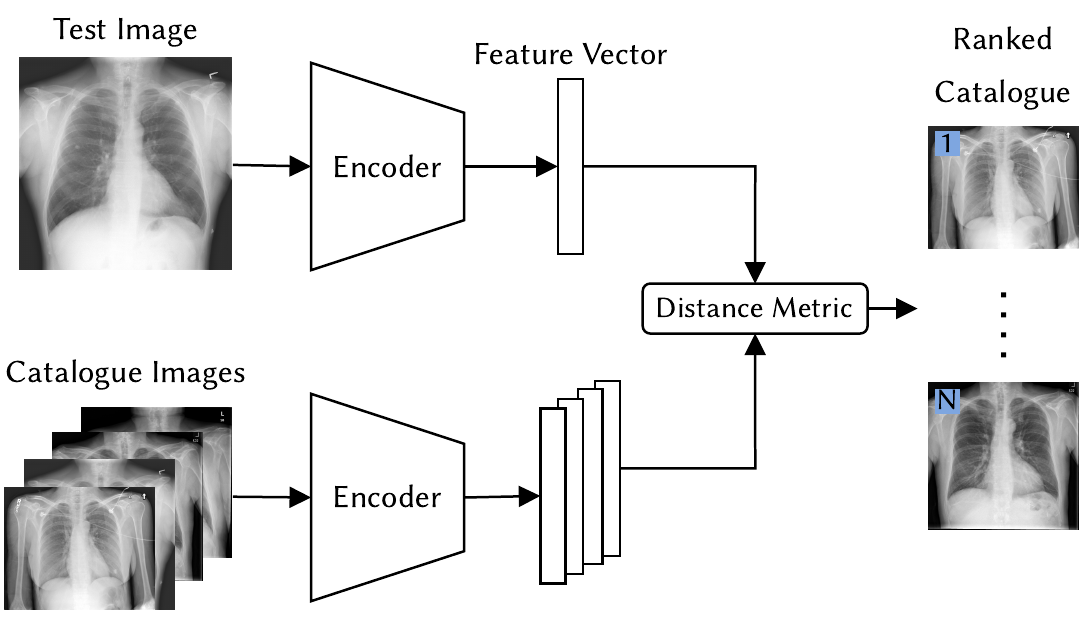}};
    \node (fig3) at (0.5\textwidth,0) 
    {\includegraphics[width=0.4\textwidth]{ 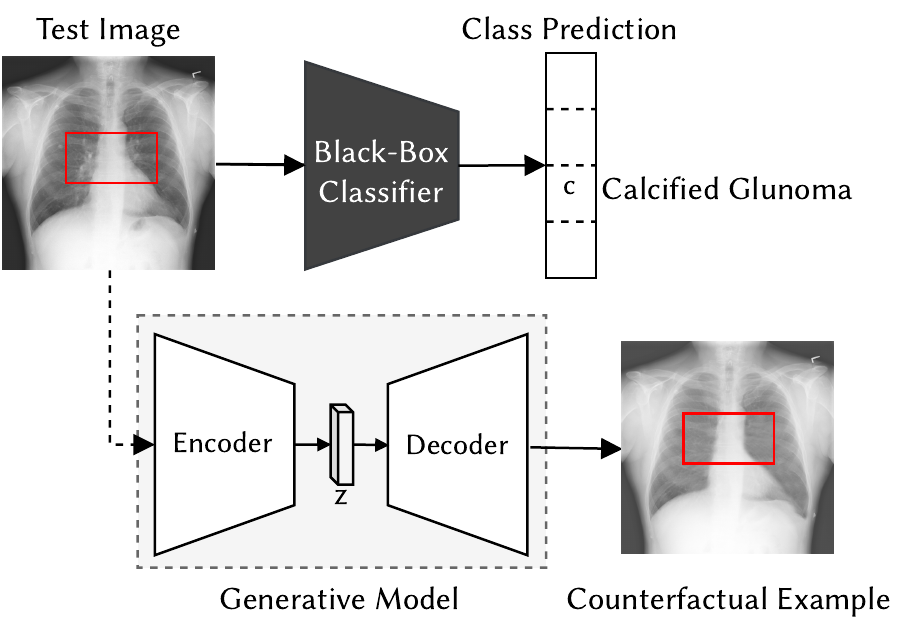}};
    
    \node at (0.03\textwidth,-2.5) {(a)};
    \node at (0.47\textwidth,-2.5) {(b)};
    
    \draw[dashed] (3.95,2.3) -- (3.95,-2.3);
    
\end{tikzpicture}
\caption{(a) \textbf{Explanation by Case-Based Reasoning}. The feature vector corresponding to the input image is compared against the feature vectors of the images in the catalogue using a distance metric, such as L2 distance. Finally, the images are retrieved from the catalogue ranked by their similarity to the input image. (b) \textbf{Explanation by Counterfactual Examples}. To explain the prediction made by a classifier, the input image is modified in a controlled way, typically by using a generative model (e.g., GAN) to shift the original class, i.e., from normal to abnormal or vice-versa. Thus, the counterfactual example intends to explain the prediction by showing that the image was classified as "abnormal" because it is not "normal", as perceived by the absence of tissue inflammation (white spots) in the generated counterfactual example.}
\label{fig:cbr+counterfactual}
\end{figure}
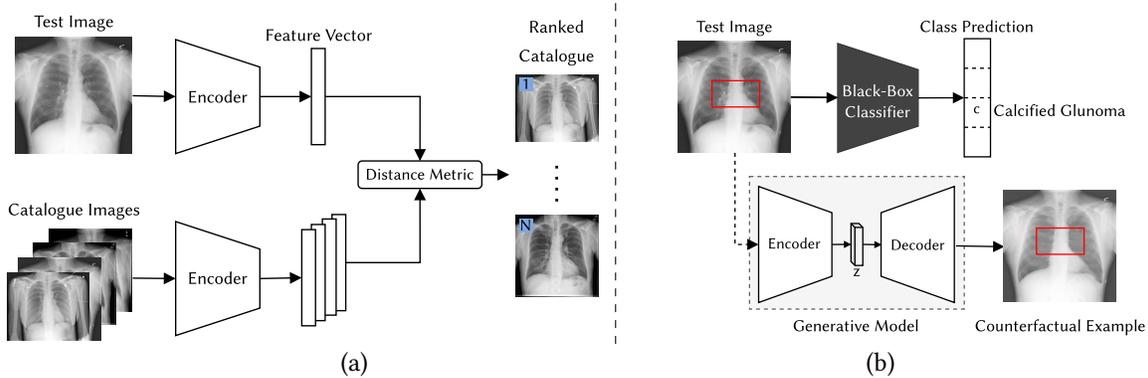

Case-Based Reasoning (CBR) and Content-Based Image Retrieval (CBIR) are example-based explanation methods that aim to search a database for visually similar entries to a specific query image. The general scheme for implementing a CBR system using DNNs is depicted in Figure \ref{fig:cbr+counterfactual}a. Although the idea of using CBIR systems in clinical settings is not novel~\cite{akgul2011content}, there has been renewed interest in CBIR approaches as a way to provide explainability to deep learning methods for medical diagnosis~\cite{chittajallu2019xai, gupta2021irtex, allegretti2021supporting}.

Recently, Barnett et al.~\cite{barnett2021interpretable} introduced a novel interpretable AI algorithm (IAIA-BL) for classifying breast masses using CBR. The model provided both a prediction of malignancy and its explanation by using known medical features (mass margins). Given an image region to analyze, the algorithm compared that region with a set of previous similar cases (image patches) using euclidean distance. The similarity score was then used to provide the mass margin scores for each medical feature, and those scores were then used to predict the malignancy score (benign or malignant). The model was trained using a fine-annotation loss penalizing activations of medically irrelevant regions on the data. The authors also introduced an interpretable evaluation metric, namely Activation Precision, to quantify the proportion of relevant information from the ``relevant region'' used to classify the mass margin regarding the radiologist annotations. The experimental results showed that the IAIA-BL achieved comparable performance to black-box models.

A different approach was presented by Tschandl et al.~\cite{tschandl2019diagnostic}, in which they compared the predictions of the ResNet-50 softmax classifier with the diagnostic accuracy obtained by using CBIR. Contrary to Barnett et al.~\cite{barnett2021interpretable}, Tschandl et al. measured the cosine similarity between two feature vectors to retrieve the most similar images to the image query. The results showed that the diagnostic accuracy obtained through CBIR is comparable to the performance of a softmax classifier leading Tschandl et al. to claim that CBIR can replace traditional softmax classifiers to improve diagnostic interpretability in a clinical workflow.

A more recent approach was introduced by Barata and Santiago~\cite{barata2021improving}, where CBIR was applied to explain the decisions of a CNN model for skin cancer diagnosis. When compared to the work of Tschandl et al., Barata and Santiago implemented an augmented category-cross entropy loss function composed of three regularization losses, namely the triplet loss, the contrastive loss, and the distillation loss. These losses encourage the model to learn a more structured feature space. The experimental results on ISIC 2018~\cite{tschandl2018ham10000} dermoscopy dataset confirmed that the combination of different loss functions lead to more structured feature spaces, which improves the performance of the classification model. Lamy et al.~\cite{lamy2019explainable} proposed an explainable CBR system with a visual interface, combining quantitative and qualitative approaches. In contrast to the above-discussed works, it used numerical data and provided a user study in which clinicians validated their approach.

The recent work of Hu et al.~\cite{hu2022x} introduced the eXplainable Medical Image Retrieval (X-MIR) approach, which explored the use of similarity-based saliency maps to explain the retrieved images visually. Concretely, they adapted the saliency map generation process for the problem of image retrieval through the use of a similarity-based formulation. These similarity-based saliency methods take as input a retrieval image and a query image for producing a saliency map highlighting the most similar regions of the retrieval image to the query image. To evaluate the quality of the generated saliency maps, the authors adapted two causal metrics, namely deletion and insertion (refer to section~\ref{subsec:evaluating_quality_visual} for a detailed description), to measure the decrease or increase in image similarity score as the retrieved image is gradually perturbed based on the important regions of its saliency map. The authors evaluated their approach on two medical datasets, namely COVIDx~\cite{wang2020covid} and ISIC 2017. They found that for both cases, the generated saliency maps focused on relevant regions when retrieved images were correct and observed the contrary when the retrieved images were incorrect. Finally, the authors pointed out for the importance of conducting user studies with clinicians to validate the utility of their approach. Silva et al.~\cite{silva2020interpretability} also explored the medical image retrieval with the addition of saliency maps to improve the class-consistency of top retrieved results while enhancing the interpretability of the whole system by accompanying the retrieval with visual explanations. 

In order to assess the effectiveness of using a CBIR system as an auxiliary tool for classifying skin lesions through dermatology images, a user-centered study was done by Sadeghi et al.~\cite{sadeghi2018users}. Sixteen non-expert users were invited to classify skin lesions images among four categories (Nevus, Seborrheic Keratosis, Basal Cell Carcinoma, and Malignant Melanoma) based on two conditions: using CBIR and without using CBIR. The results indicated that CBIR enabled users to make a significantly more accurate classification on a new skin lesion image. These findings suggest that CBIR can indeed assist clinicians understand model decisions as well as allow less experienced practitioners to improve their skills. Thus, CBIR-based systems can have a significant clinical application value as a decision-support tool to accelerate the diagnosis of pathologies.


\subsubsection{Counterfactual Explanations}

Counterfactual explanations are based on the principle that “an action on the input data will cause an outcome”~\cite{molnar2022}. The idea is to perturb the input data in a controlled way in order to reverse the final model prediction, being the modified input the counterfactual example, as illustrated in the diagram of the Figure \ref{fig:cbr+counterfactual}b. Furthermore, counterfactual explanations are deemed human-interpretable and post-hoc, meaning that they do not require access to model internals.

The work of Schutte et al.~\cite{schutte2021using} constituted a new approach in the way of interpreting the predictions made by deep learning models by using generative models to produce a sequence of images depicting the evolution of a pathology. Through the sequence of images produced, a human can understand which biomarkers triggered the prediction made by the model. Concretely, the proposed method aims to identify the optimal direction in latent space to produce a series of synthetic images with minor modifications leading to different model predictions. By observing these modified synthetic versions of the original image, it is expectable that a human can perceive the features that caused the model prediction. Experimental results on two medical image datasets showed that the proposed approach allows visualizing where the most relevant features are localized and how they contributed to the model prediction by analyzing the generated images. Moreover, this generative approach may be helpful for the identification of new biomarkers.

Kim et al.~\cite{kim2021interpretation} proposed the Counterfactual Generative Network (CGN), which is able to generate counterfactual images to explain the predictions of a pneumonia classifier from chest X-ray images. To guide the CGN towards the generation of contrastive images from query image, the prediction of the classification network was manipulated to shift the original class to the negative class. The subtraction of the counterfactual image from the input image allows the generation of attribution maps evidencing the most relevant regions to the prediction.

In the same fashion, Singla et al.~\cite{singla2021explaining} used a conditional Generative Adversarial Network (cGAN) to produce a set of counterfactual images with changed posterior probability to explain the class predictions of a chest X-ray classifier. Additionally, the context from semantic segmentation and object detection was incorporated into the loss function to preserve subtle information about the medical images in the generated counterfactual images. The validity of the generated counterfactual explanations was assessed through the use of three evaluation metrics, namely 1) Fréchet Instance Distance score to evaluate the visual quality of the counterfactual images, 2) Counterfactual Validity score to quantify the class floppiness of the counterfactual images, and 3) Foreign Object Preservation score to assess the presence of unique properties of patients in the generated explanations. Furthermore, clinical measurements, namely, cardiothoracic ratio and costophrenic recess, were adopted to demonstrate the utility of the explanations in terms of the clinical context.

{Given the recent advances in the scope of image synthesis, the use of generative diffusion probabilistic models~\cite{ho2020denoising} to produce counterfactual explanations is an interesting future research direction as it is under-explored in medical imaging. The benefit of using these models to generate counterfactual explanations is related to their ability to handle missing data and their robustness to distributional shifts~\cite{kazerouni2022diffusion}. A prominent example is the work of Sanchez et al.~\cite{sanchez2022healthy} that relied on conditional diffusion models for synthesizing healthy counterfactual examples of brain images, allowing to segment the lesion through the difference between the observed image and the healthy counterfactual.}

\begin{figure}[t]
    \centering
    \includegraphics[width=0.9\textwidth]{ 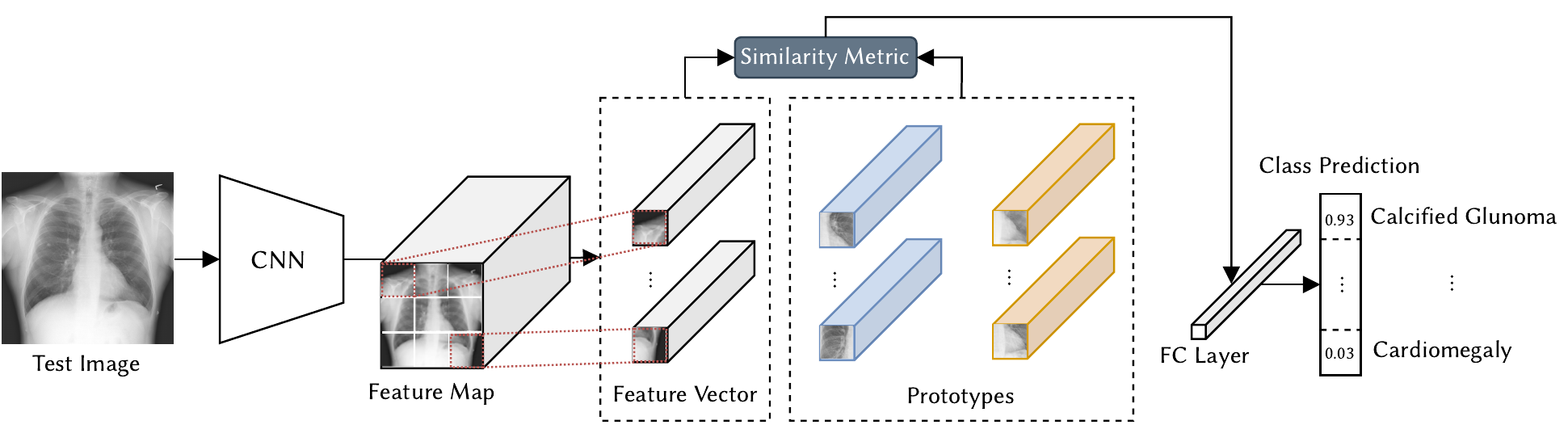}
    \caption{\textbf{Explanation by Prototypes}. During the model training, a set of prototypes are learned to visually represent a certain class (represented with blue and orange colours in the figure). In the test phase, the features extracted from the test image are compared with a set of prototypes using a similarity metric, such as cosine similarity. Then, the final class prediction is based on the similarity scores computed between the prototypes and the different parts of the input image.}
    \label{fig:prototypes}
\end{figure}

\subsubsection{Propotypes}

While most research on interpretability is still oriented towards the use of post-hoc approaches, some authors have advocated the need for devising inherently interpretable models~\cite{rudin2019stop} to obtain explanation that are indeed interpretable by humans. The learning of prototypes during the training phase of the model is a common strategy in the development of inherently interpretable models. This idea was initially explored in~\cite{chen2018looks}, where the authors incorporate a prototype layer at the end of the network~\cite{chen2018looks, barnett2021interpretable}, named ProtoPNet, for bird species recognition. The rationale behind this approach is that different parts of the image act as class-representative prototypes during training. When a new image needs to be evaluated in the testing phase, the network finds the most similar prototypes to the parts of the test image. The final class prediction is based on a score computed with the similarities between the prototypes. Figure \ref{fig:prototypes} illustrates the general pipeline for deriving a class prediction from the similarity score between different parts of the input image and a set of learned prototypes.

Based on ProtoPNet, Donnelly et al.~\cite{donnelly2021deformable} introduced the Deformable ProtoPNet. This prototypical case-based interpretable neural network provided spatially flexible deformable prototypes, i.e., prototypes that can change their relative position to detect semantically similar parts of an input image.

Despite the significance of ProtoPNet, Hoffmann et al.~\cite{hoffmann2021looks} investigated its shortcomings, and proved that ProtoPNet could be susceptible to adversarial and compression noise, and thus compromise the inner interpretability of the model. Although these limitations were not significant in the bird recognition problem, the picture changes in high-stake applications, such as the healthcare domain, where the lack of robustness can have severe implications.

Regarding the applicability of these inherently interpretable networks to medical imaging, Kim et al.~\cite{kim2021xprotonet} proposed an interpretable diagnosis framework, dubbed XProtoNet, for chest radiography to learn disease representative features within a dynamic area, using an occurrence map. Contrary to ProtoPNet, in XProtoNet the prototypes are class-representative and completely dynamic in terms of area, which is particularly important for accommodating the high variability in size of discriminative regions of medical images. To produce an appropriate occurrence map, the authors introduced two regularization terms. The L1 loss forces the occurrence area to be small enough to avoid covering irrelevant regions, and the transformation loss approximates each occurrence map with a transformed version via an affine transformation that did not change the relative location of a disease pattern. The experiments on NIH Chest X-ray dataset~\cite{wang2017chestx} confirmed that the XProtoNet surpasses the state-of-the-art models in diagnosing chest diseases from X-ray images. Later, Singh et al.~\cite{singh2021interpretable} introduced an interpretable deep learning model, named Generalized Prototypical Part Network (Gen-ProtoPNet), for detecting Covid-19 from X-ray images. Gen-ProtoPNet was inspired in the original ProtoPNet~\cite{chen2018looks} and the NP-ProtoPNet~\cite{singh2021genprotopnet}. Unlike ProtoPNet and NP-ProtoPNet that use L2 distance to calculate the similarity between prototypes, Gen-ProtoPNet used a generalized version of the L2 distance, allowing the use of prototypical parts of any dimension, i.e., squared and rectangular spatial dimensions. Furthermore, the experiments on two Covid-19 chest X-ray datasets~\cite{cohen2020covid, wang2020cord} confirmed that using prototypical parts of spatial dimensions bigger than 1 x 1 improves performance of the model, specifically when using the VGG-16 model as the feature extractor.

\subsubsection{Discussion}
Relying on example-based strategies may be a more trustworthy option since the retrieved examples are plausible and tend to contain similar findings to the input query image. However, the performance of the example-based systems can be compromised if a significant number of examples per class is unavailable. This assumption is also valid in the case of prototype-based approaches, where performance depends directly on the diversity and amount of class-representative prototypes. Regarding the counterfactual explanations, it is desirable to discover credible causal structures to create ground-truth explanations to improve further the modelling of the interventions made over the images~\cite{singla2021explaining}. 

\subsection{Explanation by Concepts}

\begin{figure}[t]
    \centering
    \includegraphics[width=0.65\textwidth]{ 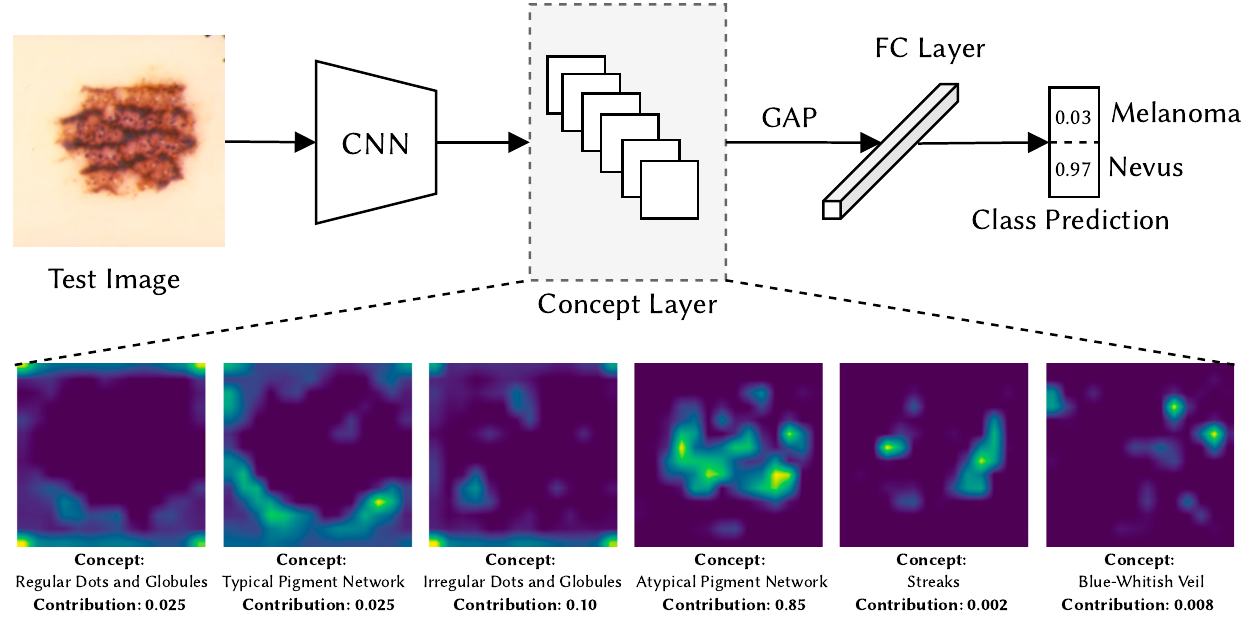}
    \caption{\textbf{Explanation by Concepts}. In the first phase, the concept layer is trained to predict the concepts associated with the input image. Then, given a test image, the model first predicts the concepts presented in the image which are subsequently processed by a fully connected layer to infer the final predictions. Simultaneously, it is possible to produce visualizations of the filters of the concept layer that highlight the relevant regions for each concept. Additionally, the contribution of each concept to the final prediction is obtained to perceive which concepts had more influence on the final decision.}
    \label{fig:concepts}
\end{figure}

The rationale behind concept-based learning approaches is using human-specified concepts as an intermediate step to derive the final predictions. This idea was used in the works of Kumar et al.~\cite{Kumar_CVPR_2009} and Lampert et al.~\cite{lampert2009learning}, with specific applications in few-shot learning approaches. In~\cite{lampert2009learning}, the proposed model first estimated a set of attributes which were subsequently used to infer the final predictions. This type of model architectures were dubbed Concept Bottleneck Models (CBM) in the work of Koh et al.~\cite{koh2020concept}. In simple terms, these models relied on an encoder-decoder paradigm, where the encoder is responsible for predicting the concepts given the raw input image, and the decoder leverages the predicted concepts by the encoder to make the final predictions. The encoder is typically a CNN model with a bottleneck layer inserted after the last convolutional layer, while the decoder can be a multi-layer perceptron to map the concepts to the final predictions. This pipeline is illustrated in Figure \ref{fig:concepts}. The idea of CBMs can be applied to any model just by inserting the bottleneck layer after the final convolutional layer. However, the main disadvantage of these methods is that annotated concepts are required. Koh et al. provided a systematic study on different ways to learn CBMs. Among the considered setup models, they concluded that the joint training is the preferred approach, which minimizes the weighted sum considering the classification loss and concept loss. The authors also stated that it is possible to intervene in the concept predictions to change the final output, {which raises the question of to what extent it is feasible to revisit, for example, 100 concepts and perceive which concept would be modified to make the correct prediction. {Later, Yuksekgonul et al.~\cite{yuksekgonul2022post} introduced Post-hoc CBM (PCBM) to address some limitations of CBM, specifically the need for concept-level annotations. The authors claim that PCBM can convert any pre-trained model into a concept bottleneck model. When concept annotations are unavailable, PCBM can leverage concept examples from other datasets and train linear binary classifiers to distinguish between examples of a single concept and negative examples. Yuksekgonul et al.~\cite{yuksekgonul2022post} identified that the problem with CBM is that they require concept-level annotations per image, which is expensive and difficult to obtain, particularly in the context of skin lesions. Concept Activation Vectors (CAVs)~\cite{kim2018interpretability} can be adopted to mitigate this by automatically predicting the presence or absence of single concepts on unseen images.}}

A different approach was introduced by Chen et al.~\cite{chen2020concept} where they proposed the Concept Whitening (CW), a module that is inserted into a neural network, and that can replace the Batch Normalization layer so that each point in the latent space has an interpretation in terms of known concepts. 
{In a similar line of research, the decision process was decomposed in a set of human-interpretable concepts, along with a visual interpretation of the spatial localization where these concepts are present in the image~\cite{wickramanayake2021comprehensible,patricio2023CVPRW}. In particular, Patrício et al.~\cite{patricio2023CVPRW} proposed an approach for enforcing the visual coherence of concept activations by using a hard attention mechanism to guide the activations of concept
filters towards the locations where to which the concept is
visually related to. This strategy has been shown to improve the visual explanation of concept-based approaches for skin lesion diagnosis.}

Ghorbani et al.~\cite{ghorbani2019towards} introduced ACE, which can automatically identify a set of high-level concepts in an unsupervised way. In a first phase, each image is segmented in multiple resolutions using the SLIC~\cite{achanta2012slic} algorithm. Then, the segments are clustered by its similarity according to the euclidean distance, which is measured in the latent space. Each group of segments represents a different concept, labelled as pseudo-concepts. Lastly, to retain only the important concepts in each group, the TCAV~\cite{kim2018interpretability} importance score was computed. The work of Fang et al.~\cite{fang2020concept} built on the rationale of the ACE. The authors proposed the Visual Concept Mining (VCM) method to explain the decisions of an infectious keratitis classifier based on human-interpretable concepts. VCM encompasses two stages: (i) the proposed Potential Concept Generator module is responsible for identifying relevant concepts based on the segmentation of image patches according to the relevant regions highlighted on the produced saliency maps; (ii) the visual concept extractor module learns the similarity and diversity among the segmented image parts and groups them according to the DeepCluster~\cite{caron2018deep} algorithm. The authors of VCM claimed that the concepts learned by their method were coherent with the medical annotations whilst being more diverse for the different classes, contrary to the ACE, which provides too broad concepts.

{A disruptive approach was proposed by Sarkar et al.~\cite{sarkar_2022_CVPR} that introduced an ante-hoc explainable model. A concept encoder on top of a backbone classification architecture is used for learning a set of human interpretable concepts providing thus an explanation for the classifier predictions. Additionally, the output of the concept encoder is passed to a decoder that reconstructs the input image, encouraging the model to capture the semantic features of the input image. Despite the method only reporting results for generic datasets, it would be interesting to extend this work for medical imaging datasets. Recently, following the philosophy of CBM, Yan et al.~\cite{yan2023towards} proposed a method to improve the trustworthiness of skin cancer diagnosis by allowing doctors to intervene in the decisions of the trained models based on their knowledge and expertise. This human-in-the-loop framework allows for discovering and removing potential confounding behaviors of the model (e.g., artifacts or bias) within the dataset during the training phase. They concluded that modifying the output of the predicted concepts lead to a more accurate model.}

\subsubsection{Discussion}
Although concept-based explanation methods are under-explored in medical imaging, they constitute a promising way of providing human-understandable explanations. The main advantage of these methods is that they are interpretable by design since the final predictions are derived from the learned concepts. However, the dependency on manual annotation of these concepts is the major limitation in concept-learning approaches. Recently, to overcome the annotation-dependency of the concepts, proposed methods built on unsupervised techniques to discover a set of pseudo-concepts related to the input image~\cite{ghorbani2019towards, fang2020concept}.  

\subsection{Other Approaches}
\label{sec:ohter_approaches}

In contrast to the above-discussed approaches, some authors have investigated alternative strategies to confer interpretability to the models, including Bayesian Neural Networks (BNN) to quantify the uncertainty associated with the model prediction or using adversarial training to improve the quality of the generated explanations.




\subsubsection{Bayesian Approaches}

Despite the success of CNN architectures, it is infeasible to quantify their uncertainty, given the deterministic nature associated with the internal parameters. Furthermore, CNNs are likely to overestimate the data when it is biased. In order to address these problems, Thiagarajan et al.~\cite{thiagarajan2021explanation} proposed the use of Bayesian CNNs (BCNN), which allow for the uncertainty estimation associated with the predictions. In particular, the uncertainty associated with the predictions of an Invasive Ductal Carcinoma (IDC) classifier on breast histopathology images was quantified. The examples characterized by a high value of uncertainty were projected into a lower-dimensional space using the t-SNE~\cite{van2008visualizing} technique to facilitate data visualization and interpretation of the test data. Furthermore, the uncertainty allowed selecting the examples requiring human evaluation, which constitutes an interesting approach in the case of problems in the medical imaging domain. Similarly, Billah and Javed~\cite{billah2022bayesian} relied on BCNNs to quantify the uncertainty associated to the predictions of a classifier model for the diagnosis of blood cancer. Recently, Gour and Jain~\cite{gour2022uncertainty} proposed the UA-ConvNet, an uncertainty-aware CNN to detect COVID-19 from chest X-ray images and provide an estimation of the model uncertainty. For this, they used Monte Carlo dropout~\cite{gal2016dropout} to obtain a probability distribution of the model prediction.

\subsubsection{Adversarial Training} 

In adversarial training, examples of the training set are augmented with adversarial perturbations at each training loop, allowing to increase the robustness of the model when provided with potential malicious examples~\cite{bai2021recent}.

The first attempt on using adversarial training to improve interpretability in a medical imaging diagnosis task was made by Margeloiu et al.~\cite{margeloiu2020improving}. They explored the use of adversarial training to improve the interpretability of CNNs, mainly when applied to diagnosing skin cancer. Specifically, the trained model was retrained from scratch using adversarial training with the Projected Gradient Descent (PGD) adversarial attack~\cite{madry2017towards}. The experiments on the dermatology dataset HAM10000~\cite{tschandl2018ham10000} showed that saliency maps of the robust model are significantly sharper and visually more coherent than those obtained from the standard trained model. 

However, further research is needed since adversarial training is under-explored in medical imaging interpretability. Specifically, applying the above-referred findings to other datasets and network architectures becomes necessary to perceive the generalization capability of the methods. Furthermore, as stated by the authors in~\cite{margeloiu2020improving}, the proposed method is not ready to be deployed in real-world scenarios due to the sensitivity of saliency methods to training noise, which can cause those methods to assign importance to artifacts available in the image (e.g., dark regions and irrelevant medical regions). Therefore, it is crucial to understand the limitations of adversarial training to improve interpretability in medical imaging diagnosis tasks.

\subsubsection{Discussion}
The uncertainty estimation associated with a classifier's predictions is helpful in the clinical workflow since clinicians can support their decisions based on the uncertainty value. Additionally, the use of adversarial training can be viewed as a method to improve the robustness of the model to adversarial attacks. As also demonstrated by Margeloiu et al.~\cite{margeloiu2020improving}, the produced explanations by adversarially trained models seem to be more coherent and sharpening. Despite the under-exploitation of these strategies in medical imaging, these preliminary findings may encourage the emergence of methods adopting these alternative strategies as an additional layer to improve the models' reliability and robustness.

\section{Evaluation Metrics}
\label{sec:evaluation_metrics}

Depending on the type of explanation modality (visual or textual), there are different ways to assess the quality of the generated explanations. We divide the evaluation metrics used in the literature into two categories: (i) evaluation metrics to assess the quality of visual explanations; and (ii) evaluation metrics for measuring the quality of textual explanations.
\subsection{Evaluating the Quality of Visual Explanations}
\label{subsec:evaluating_quality_visual}

Evaluating the quality of model explanations remains an active area of research. A common approach for evaluating the model interpretability, specifically when applied to the medical domain, is to request an expert opinion from the clinicians and radiologists. However, this evaluation approach is time-consuming and depends on the level of experience of the clinicians ~\cite{gale2018producing, tsiknakis2020interpretable}.

Therefore, there has been attempts to propose evaluation metrics capable of objectively assessing the quality of the explanations. Samek et al.~\cite{samek2016evaluating} were precursors in contributing to the question of how to objectively evaluate the quality of heatmaps by introducing the area over the Most Relevant First (MoRF) perturbation curve (AOPC) measure. This measure is based on a region perturbation strategy that iteratively removes information from some regions of the input image according to its relevance to the class, allowing to perceive the performance decay of the model. The conducted experiments showed that a large AOPC value denotes high model sensitivity to the perturbations, indicating that the heatmap is actually informative.

Inspired by the work in~\cite{samek2016evaluating}, Petsiuk et al.~\cite{petsiuk2018rise} proposed two causal metrics, namely deletion and insertion, to evaluate the produced explanations for a black-box model. The deletion metric measures the degradation of the class probability, as important pixels of the image, derived from the saliency map, are removed. On the other hand, the insertion metric intends to measure the increase of the class probability, as pixels are inserted based on the generated saliency map.

Later, Hooker et al.~\cite{hooker2019benchmark} argued that the modification-based metrics introduced by Petsiuk et al.~\cite{petsiuk2018rise} might not capture the actual reasoning behind the model's degradation since this degradation could be due to artefacts introduced by the values used to replace the removed pixels. Thus, the authors proposed the RemOve And Retrain (ROAR) to evaluate interpretability methods by verifying how the accuracy of a retrained model degrades as important features are removed. The most important features are removed in certain regions of the image with a fixed uninformative value for each interpretability method. The main drawback of this metric is the required retraining of the model, which is computationally expensive. 

Recently, Rio-Torto et al.~\cite{rio2020understanding} proposed the POMPOM (Percentage of Meaningful Pixels Outside the Mask) metric, which determines the number of meaningful pixels outside the region of interest in relation to the total number of pixels, to evaluate the quality of a given explanation. Similarly, Barnett et al.~\cite{barnett2021interpretable} introduced the Activation Precision evaluation metric, to quantify the proportion of relevant information from the “relevant region” used to classify the mass margin regarding the radiologist annotations. Despite the relevance of both metrics, they require manual annotations of the masks, which is time-consuming and may be difficult to obtain for some medical image datasets. 

In spite of the valuable contribution of these proposed evaluation metrics, we believe that a new evaluation method for model interpretability can be developed with the aid of the Bayesian Neural Networks (BNN) characteristics. Considering that the weights of the BNN follow a probability distribution, we can sample different ``models'' from the posterior distribution and generate an arbitrary number of explanations for a given example \cite{bykov2021explaining}. Then, using intersection or union operations over those generated explanations seems to be an interesting direction to estimate whether most of the explanations highlight the same Region Of Interest (ROI).


\subsection{Evaluating the Quality of Textual Explanations}

In this category, the metrics are used to measure the quality of the generated text, and are originated from generic NLP tasks. The most used metrics in the reviewed papers were BLEU~\cite{papineni2002bleu}, ROUGE-L~\cite{lin2004rouge}, METEOR~\cite{lavie2007meteor} and CIDEr~\cite{vedantam2015cider}.

BLEU (Bilingual Evaluation Understudy) score is the most common evaluation metric in NLP. In simple terms, it compares \textit{n}-gram matches between the generated sentence (also known as candidate sentence) and the ground-truth sentence (also known as reference sentence), expressed in modified precision\footnote{Takes into consideration the maximum frequency of each n-gram in the reference sentence (clipped count). The modified precision is then calculated by summing the clipped counts of the candidate sentence divided by the total number of candidate n-grams.} for each \textit{n}-gram. The BLEU metric has \textit{N} variations (BLEU-N), typically \textit{N} $\in \{1, 2, 3, 4\}$, with respect to the considered \textit{n}-grams. The value of BLEU score ranges from 0 to 1, which means that the closer to 1, the better the translation. Formally, BLEU score is calculated according to the following formulation:

\begin{equation}
    BLEU = BP . \exp{\sum_{n=1}^{N} w_n \log{p_n}},
\end{equation}

where $p_n$ is the modified precision for $n$-gram, $w_n$ is a weight, ranging from $0$ and $1$, and $\sum_{n=1}^{N} w_n = 1$, i.e., if $N=4$, $w_n = 1/N$, and BP is the brevity penalty that penalizes short generated sentences, denoted as:

\begin{equation}
    BP = \left\{\begin{matrix}
 1 & \textrm{if} \; \; c > r\\ 
 \exp(1 - \frac{1}{c}) & \textrm{if} \; \; c \leq r 
\end{matrix}\right.,
\end{equation}

where $c$ is the length of candidate translation, and $r$ is the reference corpus length.

The ROUGE (Recall-Oriented Understudy for Gisting Evaluation) metric is actually a set of metrics. We focus on the ROUGE-L variant as it was the prevalent metric in the most of the reviewed methods. ROUGE-L measures the Longest Common Subsequence (LCS) between the generated sentence and the ground-truth sentence both in terms of precision and recall. This means that if both sentences share a long sub-sentence, the similarity between the two sentences is expected to be high. The final value of ROUGE-L is given in F1 score, as formally described bellow (Eq. \ref{eq:rougel}):



\begin{equation}
\label{eq:rougel}
    ROUGE-L_{F1} = 2 \cdot \frac{Precision \cdot Recall}{Precision + Recall},
\end{equation}

where $Precision = \frac{LCS(c, r)}{m}$ and $Recall = \frac{LCS(c, r)}{n}$, with $LCS(c, r)$ denoting the Longest Common Subsequence between the candidate sentence (c) and the reference sentence (r), $m$ is the number of $n$-grams in the reference sentence, and $n$ the number of $n$-grams in the candidate sentence.

Differently from the two above-discussed metrics, METEOR (Metric for Evaluation of Translation with Explicit Ordering) gives attention to the position of the words in the sentence by including a chunk penalty that weights the final score.

\begin{equation}
    METEOR = F_{mean} \cdot (1 - Penalty) 
\end{equation}



where $Penalty = 0.5 \cdot (\frac{m}{n})$, where $m$ is the number of chunks and $n$ is the total number of unigram matches, and $F_{mean} = \frac{10PR}{R+9P}$.

Alternatively, CIDEr (Consensus-Based Image Description Evaluation) is an evaluation metric that uses the Term Frequency Inverse Document Frequency (TF-IDF)~\cite{robertson2004understanding} for weighting each $n$-gram. The intuition behind CIDEr is that $n$-grams that frequently appear in the reference sentences are less likely to be informative, and hence they have a lower weight using the IDF term. CIDEr$_n$, $n = \{1, 2, 3, 4\}$, is the average cosine similarity between the candidate sentence and the reference sentences, considering both precision and recall. 

TF-IDF weighting $g_k(s_{ij})$ for each $n$-gram $w_k$ is computed as follows:

\begin{equation}
    g_k(s_{ij}) = \frac{h_k(s_{ij})}{\sum_{w_l \in \omega} h_l(s_{ij})} \log (\frac{\left | I \right |}{\sum_{I_p \in I}} \min (1, \sum_q h_k (s_{pq}))),
\end{equation}

where $h_k(s_{ij})$ is the number of times an $n$-gram occurs in a reference sentence $s_{ij}$, and $h_k(c_i)$ for the candidate sentence $c_i$, $\omega$ is the vocabulary of all $n$-grams, and $I$ is the set of all images.

\begin{equation}
    CIDEr_n(c_i, S_i) = \frac{1}{m} \sum_j \frac{g^n(c_i)  \cdot g^n(s_{ij})}{\|g^n(c_i)\|\|g^n(s_{ij})\|}.
\end{equation}

The final CIDEr score is the weighted average of CIDEr$_n$:

\begin{equation}
    CIDEr(c_i, S_i) = \sum_{n=1}^{N} w_k  CIDEr_n(c_i, S_i),
\end{equation}

where $w_n = 1 / N, N=4$.


\section{Performance Comparison}
\label{sec:comparison}

In the previous sections, we reviewed several works focused on providing explanations to the output of automated medical diagnosis. At the end of the review, an important question arises: ``\textit{What is the best approach?}''. Unfortunately, in many cases, there is no trivial answer, since most methods adopt different evaluation metrics making performance comparison with other competing methods unfeasible. To add up to the question, we compare the performance of some of the methods reviewed. In order to find common ground for a fair comparison of the methods, only those methods that considered the same dataset for evaluation purposes were selected. As presented in table \ref{tab:methods}, IU Chest X-ray~\cite{demner2016preparing} is the most used dataset among the reviewed methods. As such, we verified whether the methods that were evaluated on the IU Chest X-ray~\cite{demner2016preparing} used the same evaluation metrics. 
{Additionally, and since MIMIC-CXR~\cite{johnson2019mimic} dataset provides an official training-test partition, we include some methods that report results on this dataset under the same evaluation metrics.} In this way, a comparison of the performance between these methods was carried out. {Table \ref{tab:methods} conveys the results in terms of a set of NLP evaluation metrics (BLEU score, ROUGE, METEOR, and CIDEr) for each of the selected methods.}


\begin{table*}[t]
  \caption{Performance of the selected methods that generate textual explanations for interpreting the decision of a classifier. Spaces marked with a ``-'' mean no value is available for the respective metric. $^1$ Results were taken from the work of Liu et al.~\cite{liu2019clinically}. {The methods evaluated on the IU Chest X-ray dataset are marked with a symbol {($\star$, $\dagger$, $\circ$, $\ddagger$)}, meaning that the methods with each symbol used the same training-validation-test split.}}
  \label{tab:methods}
  \resizebox{0.9\textwidth}{!}{%
  \begin{tabular}{clcccccccc}
    \toprule
    &\textbf{Model} & \textbf{Year} & \textbf{BLEU-1} & \textbf{BLEU-2} & \textbf{BLEU-3} & \textbf{BLEU-4} & \textbf{ROUGE-L} & \textbf{METEOR} & \textbf{CIDEr} \\
    \midrule
    &\multicolumn{9}{c}{\textit{IU Chest X-Ray}} \\
    \midrule
    \multirow{4}{*}{\rotatebox[origin=c]{90}{\textit{\makecell{Find. +\\Impress.}}}}
    &Jing et al.~\cite{jing2017automatic}$^\dagger$ & 2017 & ${0.517}$ & ${0.386}$ & $\textbf{0.306}$ & $\textbf{0.247}$ & ${0.447}$ & $\textbf{0.217}$ & $0.327$ \\
    &Singh et al.~\cite{singh2019chest}$^\circ$ & 2019 & $0.374$ & $0.224$ & $0.153$ & $0.110$ & $0.308$ & $0.164$ & ${0.360}$ \\
    &HRNN~\cite{yin2019automatic}$^\dagger$ & 2019 & $0.445$ & $0.292$ & $0.201$ & $0.154$ & $0.344$ & $0.175$ & $0.342$ \\
    &{Selivanov et al.~\cite{selivanov2023medical}$^\dagger$} & 2023 & $\textbf{0.520}$ & $\textbf{0.390}$ & $0.296$ & $0.235$ & $\textbf{0.450}$ & $-$ & $\textbf{0.701}$ \\
    \midrule
    \multirow{10}{*}{\rotatebox[origin=c]{90}{\textit{Findings}}}
    &HRGR-Agent~\cite{li2018hybrid}$^\star$ & 2018 & $0.438$ & $0.298$ & $0.208$ & $0.151$ & $0.322$ & $-$ & $0.343$ \\
    &TieNet$^1$~\cite{wang2018tienet}$^\star$ & 2018 & $0.330$ & $0.194$ & $0.124$ & $0.081$ & $0.311$ & $-$ & $1.334$\\
    &Liu et al.~\cite{liu2019clinically}$^\star$ & 2019 & $0.369$ & $0.246$ & $0.171$ & $0.115$ & $0.359$ & $-$ & $\textbf{1.490}$ \\ 
    &R2Gen~\cite{chen2020generating}$^\star$ & 2020 & ${0.470}$ & ${0.304}$ & ${0.219}$ & ${0.165}$ & ${0.371}$ & ${0.187}$ & $-$ \\
    &{PPKED~\cite{liu2021exploring}$^\star$} & 2021 & $0.483$ & $0.315$ & $0.224$ & $0.168$ & $0.376$ & $-$ & $0.351$ \\
    &{CA~\cite{liu2021ACL}$^\star$} & 2021 & $0.492$ & $0.314$ & $0.222$ & $0.169$ & $0.381$ & $0.193$ & $-$ \\
    &{ICT~\cite{zhang2023novel}$^\star$} & 2023 & ${0.503}$ & $\textbf{0.341}$ & $\textbf{0.246}$ & ${0.186}$ & ${0.390}$ & $\textbf{0.208}$ & $-$ \\
    &{METransformer~\cite{wang2023metransformer}$^\star$} & 2023 & $0.483$ & $0.322$ & $0.228$ & $0.172$ & $0.380$ & $0.192$ & $0.435$\\
    &{VLCI~\cite{chen2023visual}$^\star$} & 2023 & $\textbf{0.505}$ & $0.334$ & $0.245$ & $\textbf{0.189}$ & $0.397$ & $0.204$ & $0.456$ \\
    &{Yang et al.~\cite{yang2023radiology}$^\star$} & 2023 & ${0.497}$ & $0.319$ & $0.230$ & $0.174$ & $\textbf{0.399}$ & $-$ & $0.407$\\
    \midrule
    &\multicolumn{9}{c}{\textit{MIMIC-CXR}} \\
    \midrule
    &{TieNet$^1$~\cite{wang2018tienet}$^\ddagger$} & 2018 & $0.332$ & $0.212$ & $0.142$ & $0.095$ & $0.296$ & $-$ & $1.004$ \\
    &{Liu et al.~\cite{liu2019clinically}$^\ddagger$} & 2019 & $0.352$ & ${0.223}$ & ${0.153}$ & ${0.104}$ & ${0.307}$ & $-$ & ${1.153}$ \\
    &R2Gen~\cite{chen2020generating} & 2020 & ${0.353}$ & $0.218$ & $0.145$ & $0.103$ & $0.277$ & $0.142$ & $-$ \\
    &{CA~\cite{liu2021ACL} }& 2021 & $0.350$ & $0.219$ & $0.152$ & $0.109$ & $0.283$ & ${0.151}$ & $-$ \\
    &{PPKED~\cite{liu2021exploring}} & 2021 & $0.360$ & $0.224$ & $0.149$ & $0.106$ & $0.284$ & $0.149$ & $-$ \\
    &{MSAT~\cite{wang2022medical}} & 2022 & $0.373$ & $0.235$ & $0.162$ & $0.120$ & $0.282$ & $0.143$ & $0.299$ \\
    &{ICT~\cite{zhang2023novel}} & 2023 & ${0.376}$ & ${0.233}$ & ${0.157}$ & ${0.113}$ & $0.276$ & $0.144$ & $-$ \\
    &{METransformer~\cite{wang2023metransformer}} & 2023 & $0.386$ & ${0.250}$ & ${0.169}$ & ${0.124}$ & $0.291$ & $\textbf{0.152}$ & $0.362$\\
    &{Yang et al.~\cite{yang2023radiology}} & 2023 & $0.386$ & $0.237$ & $0.157$ & $0.111$ & $0.274$ & $-$ & $0.111$\\
    &{VLCI~\cite{chen2023visual}} & 2023 & $0.400$ & $0.245$ & $0.165$ & $0.119$ & $0.280$ & $0.150$ & $0.190$ \\
    &{Selivanov et al.~\cite{selivanov2023medical}} & 2023 & $\textbf{0.725}$ & $\textbf{0.626}$ & $\textbf{0.505}$ & $\textbf{0.418}$ & $\textbf{0.480}$ & $-$ & $\textbf{1.989}$ \\

    \bottomrule
  \end{tabular}%
  }
\end{table*}

\subsection{Results Discussion}

It is worth noticing that all the presented results in Table \ref{tab:methods} were taken from the original paper of each method, except for the TieNet~\cite{wang2018tienet} method, whose results were taken from the work of Liu et al.~\cite{liu2019clinically}, since in the original paper the authors only provide results in ChestX-ray14 using BLEU, METEOR, and ROUGE-L.

Regarding the methods that considered the ``findings+impression'' section from the radiology report to generate the free-text report, and as evidenced by the results in Table \ref{tab:methods}, the model proposed by Selivanov et al.~\cite{selivanov2023medical} proved to be more accurate in terms of BLUE-1, BLEU-2 and ROUGE-L. The preprocessing and squeezing approaches used for clinical records jointly with the combination of two large language models (Show-Attend-Tell (SAT)~\cite{xu2015show} and Generative Pretrained Transformer (GPT-3)~\cite{brown2020language}) can explain the performance improvement. Moreover, generated reports are accompanied by 2D heatmaps that localize each pathology on the input scans. {Conversely, the method of Jing et al.~\cite{jing2017automatic} demonstrates superior performance in terms of BLEU-3, BLEU-4 and METEOR, which can be explained by the co-attention mechanism adopted by the authors. As the method of Singh et al.~\cite{singh2019chest} does not follow the same data partition, it is not strictly comparable to other methods.}



{Among the selected methods that only consider the ``findings'' section, ICT~\cite{zhang2023novel} demonstrates superior performance on BLEU-2, BLEU-3 and METEOR metrics. Their transformer-based model incorporates two modules responsible for capturing inter-intra features of medical reports as auxiliary information and subsequently calibrating the report generation process by integrating that information. This combination results in a performance boost in the report generation model and improves the quality of medical diagnosis. In contrast, with regard to the BLEU-1 score, VLCI~\cite{chen2023visual} exhibits a superior performance, comparable to ICT~\cite{zhang2023novel}, which could be justified by the cross-modal causal intervention strategy employed by the authors to mitigate spurious correlations from visual and linguistic confounders. Furthermore, the model proposed by Liu et al.~\cite{liu2019clinically} adopted a fine-tuning procedure that uses reinforcement learning via CIDEr to ensure more coherent report generation, which could justify the performance in the CIDEr metric.}


{Regarding the methods that report results for the MIMIC-CXR dataset, it is worth noting that only TieNet~\cite{wang2018tienet} and the approach by Liu et al. [82] have followed a distinct data partition strategy. Consequently, these methods are included here primarily for reference, as their divergent data partitioning makes direct comparisons with other methods infeasible.
On the contrary, the model proposed by Selivanov et al.~\cite{selivanov2023medical} distinguishes itself from the others by exhibiting superior performance across all metrics except for METEOR (which was not reported by the authors). Their approach leveraged the capabilities of two large language models, SAT and GPT-3, trained on large text corpora. This fusion of language models significantly improved standard text generation scores, as shown in Table \ref{tab:methods}.}

Overall, it is noticeable in Table \ref{tab:methods} that the use of transformer-based models~\cite{liu2021exploring,liu2021ACL,wang2022medical,wang2023metransformer, yang2023radiology} with additional mechanisms to capture complex and relevant features proved to be effective in improving the performance of the generated reports, as observed in the results obtained in the MIMIC-CXR dataset.

\section{General Discussion}
\label{sec:discussion}

Although XAI is a relatively recent research field, its constant growth is undeniable, with applications in many areas, particularly in the medical domain. However, despite the advances and the efforts made toward developing interpretable deep learning-based models for medical imaging, there are open issues that require more research and advances in this growing field. This section identifies open challenges in the literature and potential research paths to further improve the trustworthiness of provided explanations and foster the adoption of deep learning-based systems into clinical routine.

Based on reviewed literature, it can be concluded that the go-to method for model interpretation in medical imaging is producing saliency maps, using classical techniques, such as Grad-CAM, Integrated Gradients, or LRP. However, as evidenced by some authors, saliency maps can be unreliable and fragile~\cite{rudin2019stop, adebayo2018sanity}, as they often highlight irrelevant regions in the images. In addition, it is frequent that very similar explanations are given for different classes, and often none of them are useful explanations~\cite{rudin2019stop}. {Thus, the development of inherently interpretable models has been a line of research with promising results in the medical imaging domain. Although these methods remain largely unexplored in medical imaging, future research will undoubtedly be devoted to develop inherently interpretable models. These models have the primary benefit of providing their own explanations, which contributes to their transparency and fidelity, increasing the chances of being adopted into the clinical routine.} 

{On the other hand, different end-users could have different backgrounds and preferences at interpreting the explanation, which can generate some contradictory opinions. This fostered the use of textual explanations, which are preferred over visual explanations by some authors~\cite{gale2018producing} since they are inherently understandable by humans~\cite{vandervelden2021explainable}. Since then, methods that generate textual descriptions for explaining a prediction and multimodal methods that combine visual and textual explanations have emerged. However, generating free-text reports is deemed a challenging task since the radiologist reports are technically structured, and the most used language models based on RNNs have some limitations in generating long texts, as stated by Pascanu et al.~\cite{pascanu2013difficulty}.}

As an alternative to text-based explanations, the use of example-based explanations was proposed, since this explanation modality is directly linked to how humans try to explain something to the other humans. This way, some example-based approaches have emerged with promising results that were even comparable to the performance of standard classifiers. These example-based methods include CBR approaches, prototype-based and concept-based strategies. Recently, Schutte et al.~\cite{schutte2021using} introduced a disruptive approach that strives to generate synthetic examples to explain a model decision, as discussed in section \ref{sec:ohter_approaches}. The possible limitations of using example-based methods are related to the availability of a considerable amount of data covering all the classes in a balanced way, without forgetting the sensibility of these methods to adversarial attacks, even though this can be prevented by using adversarial training~\cite{margeloiu2020improving}.


Alternatively, other methods have emerged as candidates for explaining the decision of a model. The adoption of Bayesian Neural Networks to estimate or quantify the uncertainty regarding the model predictions might be an interesting option, although few works attempted to prove its effectiveness. 



Future research in medical image interpretability can also include the use of vision transformers (ViTs)~\cite{dosovitskiy2020image}. According to~\cite{matsoukas2021time}, vision transformers proved to be comparable with CNN in terms of performance (accuracy) in the medical classification tasks. Furthermore, ViTs have the benefit of providing a type of built-in saliency maps that are used to better understand the model’s decisions. 

Regarding medical image datasets, existing publicly available datasets for medical image captioning are limited in number and there is need to generate more large size datasets. Moreover, most of the existing datasets has focused only on few anatomical parts of body, such as chest, while ignoring other important parts like breast and brain~\cite{AYESHA2021107856}.




\subsection{Challenges and Future Research Trends}

Despite the rapid pace of advances in the medical imaging and deep learning, there are problems that remain without a definitive solution.

\begin{itemize}
    \item \textbf{Small datasets}: The collection of medical data depends on multiple entities and background bureaucracies. Nevertheless, the main issue is related to the availability of the physicians in annotating a vast amount of data, that is time-consuming and costly. {This is even more critical in the XAI field, where additional annotations are required (e.g., concepts, textual descriptions). For this reason, interpretability-compliant medical datasets have a lower representativeness of the classes, resulting in poor generalizability and applicability of the developed methods to real-world scenarios}. To surpass these constraints, distinct data augmentation techniques have emerged as an alternative for collecting new data. Recently, Wickramanayake et al. ~\cite{wickramanayake2021explanation} proposed the BRACE framework to augment the dataset based on concept-based explanations from model decisions, which can help to discover the samples in the under-represented regions in the training set. Furthermore, Wickramanayake et al. introduced a utility function to select the images in the under-representation regions and concepts that caused the misclassification. The images with a high utility score are selected to incorporate the training set.  On the other hand, the use of generative approaches to perform data augmentation in a controlled way might be an interesting research direction.
    \item \textbf{Insufficient labelled data}: Although most works rely on the supervised learning paradigm, it is often not the best choice when working in the medical domain, since the process of label annotation is time-consuming and costly for large-scale datasets, especially in domains such as digital pathology where the manual annotation is subject to inter- and intra-observer variability~\cite{yu2021convolutional}. Transfer learning was adopted in most works to address these issues, but this technique is not completely effective in the medical domain, since the original models were trained in images belonging to standard object detection datasets (e.g., ImageNet), which do not share the same patterns of medical imagery. This way, Self-Supervised Learning (SSL) has emerged to tackle these challenges, allowing the network to learn visual meaningful feature representations without the need of annotated data~\cite{chowdhury2021applying}. Besides its effectiveness in dealing with scarce labelled data, it confers robustness to the model, rendering it more resistant to adversarial attacks. In medical imaging, the use of SSL seems to be a promising research direction due to the characteristics of medical datasets. Furthermore, contrastive learning approaches have achieved impressive results, due to the contrastive loss that encourages the network to learn high-level features that occur in images across multiple views, which are created through the use of geometric transformation such as random cropping, color distortion, gaussian blur. For a comprehensive overview of the state-of-the-art of SSL with a particular focus on medical domain we refer the reader to~\cite{chowdhury2021applying}. {The inherently interpretable models, specifically the concept bottleneck models, require the annotation of concepts for each class or image. The presented datasets in Table \ref{tab:datasets} show that the majority does not include this type of annotation, hampering the rapid employment of the concept bottleneck models. Furthermore, despite the significance of the existing methods that emerged to surpass these issues (CAV~\cite{kim2018interpretability}), annotations regarding clinical concepts remain necessary. This could be solved with closer cooperation between clinicians and the AI community. As discussed in Section \ref{subsec:discussion_datasets}, appropriate annotations in medical imaging datasets are needed to ensure a quick development of interpretability methods for medical diagnosis.}
    \item \textbf{Qualitative assessment of the explanations}: The automated evaluation of the explanations provided by interpretability methods remains an open challenge. As previously discussed in section \ref{subsec:evaluating_quality_visual}, the most adopted method for evaluating the explanations in the context of the medical domain is to resort to the clinician's expertise. However, considering variability in experts opinions~\cite{tonekaboni2019clinicians}, this strategy is particularly biased and subjective. On the other hand, the existing strategies for objectively measuring the quality of visual explanations are still dependent on manual annotations of relevant regions~\cite{rio2020understanding}, or iterative model retraining (ROAR~\cite{hooker2019benchmark}). For these reasons, we believe that the design of objective metrics for assessing the quality of the model explanations will be one of the important research trends on the topic of XAI.
    \item \textbf{Report generation in medical imaging}: Text-based explanations are usually obtained using RNN-based approaches through the generation of words forming a sentence. Nevertheless, RNN-based approaches have some limitations in generating long text reports~\cite{pascanu2013difficulty}. Consequently, the use of Transformers for the automatic generation of radiology reports was adopted in an attempt to overcome the limitations of traditional RNNs, namely the problem of vanishing gradients. The self-attention mechanism of the Transformer architecture allows for the learning of contextual relationships between the words that constitute the sequence. In addition, Transformer-based networks can be trained faster than traditional RNNs as they allow for simultaneous processing of sequential data. {With the rise of foundation models, such as Generative Pretrained Transformers~\cite{selivanov2023medical}, the limitations encountered in previous methods have been alleviated, specifically the coherency of the generated texts.} On the other hand, using concept-based approaches as a transition bridge between free-text report generation and concept-based explanations may be an exciting future research direction. Instead of trying to generate free-text reports, which is challenging, having a set of concepts that are sufficient to describe the phenomenon depicted in the image can support clinicians in writing a complete report. 
    \item {\textbf{Deployment in clinical practice}: The implementation of XAI methods in clinical practice requires rigorous validation to ensure their safety, effectiveness, and reliability, which can be challenging due to the complex and dynamic nature of clinical environments. Additionally, the field of medical imaging is subject to rigorous regulations, and the development and deployment of XAI methods must comply with regulatory and legal requirements~\cite{goodman2017european}, such as FDA approvals~\cite{benjamens2020state}, data privacy regulations, and liability concerns.}
\end{itemize}





\section{Conclusions}
\label{sec:conclusions}

This paper reviewed the advances on explainable deep learning applied to medical imaging diagnosis. First we introduced a comparative analysis between the existing surveys on the topic, where the major conclusions and weaknesses of the each were highlighted.
Then, the most prominent XAI methods were briefly described to provide the readers with fundamental concepts of the field, necessary to the discussion of the recent advances on the medical imaging domain. Additionally, several frameworks that implement XAI methods were presented and a brief discussion of the existing medical imaging datasets was drawn. After this, we comprehensively reviewed the works focused on explaining the decision process of deep learning applied to medical imaging. The works were grouped according to the explanation modality comprising explanations by feature attribution, explanations by text, explanation by examples and explanations by concepts. Contrary to other surveys on the topic, we focused this review on inherently interpretable models over post-hoc approaches, which has been recently considered a future research direction on deep learning interpretability. The discussion of the adopted evaluation metrics used in the literature was also carried out, where we described the existing metrics to assess the quality of visual explanations and the commonly NLP metrics to evaluate the quality of the generated textual explanations. Additionally, a comparison of the performance of a set of prominent XAI methods was performed based on the dataset used and the evaluation metrics adopted. Finally, the discussion and future outlook in XAI for medical diagnosis were addressed where we identified open challenges in the literature and potential research avenues to improve the trustworthiness of provided explanations and foment the adoption of deep learning-based systems into clinical routine. To conclude, we believe that this survey will be helpful to the XAI community, particularly to the medical imaging field, as an entry point to guide the research and the future advances in the topic of XAI.

\section*{Acknowledgements}
This work was funded by the Portuguese Foundation for Science and Technology (FCT) under the PhD grant ``2022.11566.BD'', and supported by NOVA LINCS (UIDB/04516/2020) with the financial support of FCT.IP.




\bibliographystyle{ACM-Reference-Format}
\bibliography{sample-base}

\clearpage
\appendix

\section{Appendix}
\label{sec:suppl}

\subsection{Intepretability Frameworks}
\label{appx:int_frameworks}

The increasing interest in the interpretability of machine learning fostered the development of frameworks implementing classical XAI methods. The LRP Toolbox~\cite{lrp_toolbox} was launched in 2016 and provides the implementation of the LRP~\cite{bach2015pixel} algorithm for artificial neural networks supporting Matlab and Python. Additionally, the LRP toolbox released an extension to be compatible with the Caffe Deep Learning framework. DeepExplain~\cite{ancona2017towards} is a framework that implements perturbation and gradient-based attribution methods, including Saliency Maps~\cite{simonyan2014deep}, Gradient * Input~\cite{pmlr-v70-shrikumar17a}, Integrated Gradients (IG)~\cite{sundararajan2017axiomatic}, DeepLIFT~\cite{pmlr-v70-shrikumar17a}, $\epsilon$-LRP~\cite{bach2015pixel}, and DeConvNet~\cite{zeiler2014visualizing}. It also supports Tensorflow as well as Keras with Tensorflow backend. Alternatively, iNNvestigate~\cite{JMLR:v20:18-540} is a more complete toolbox that provides implementations for SmoothGrad~\cite{smilkov2017smoothgrad}, DeConvNet~\cite{zeiler2014visualizing}, Guided-BackProp~\cite{springenberg2014striving}, PatternNet~\cite{kindermans2017learning}, Input * Gradients~\cite{pmlr-v70-shrikumar17a}, DeepTaylor~\cite{montavon2017explaining}, PatternAttribution~\cite{kindermans2017learning}, LRP~\cite{bach2015pixel} and Integrated Gradients (IG)~\cite{sundararajan2017axiomatic}. It also supports Tensorflow and Keras.

With regard to PyTorch frameworks, TorchRay~\cite{fong2019understanding} implements several visualization methods, namely Gradient~\cite{simonyan2014deep}, Guided-BackProp~\cite{springenberg2014striving}, Grad-CAM~\cite{selvaraju2017grad} and RISE~\cite{petsiuk2018rise}. TorchRay is considered research-oriented, since it provides code for reproducing results that appear in several papers. Captum~\cite{captum} is a PyTorch library that provides state-of-the-art algorithms for model interpretability and understanding. It contains general purpose implementations of Integrated Gradients~\cite{sundararajan2017axiomatic}, SmoothGrad~\cite{smilkov2017smoothgrad}, VarGrad~\cite{richter2020vargrad} and others for PyTorch models. Table \ref{tab:software} summarizes the aforementioned frameworks alongside the supported XAI methods. Additionally, we refer the reader to the work of Darias et al.~\cite{darias2021systematic} in which some other model-agnostic XAI libraries were approached.

\begin{table}[h!]
    \centering
    \caption{Publicly available interpretability frameworks.}
    \resizebox{0.9\textwidth}{!}{%
    \begin{tabular}{lccc}
    \toprule
         \textbf{Framework} & \textbf{Year} & \textbf{Methods} & {\textbf{\makecell{Supported DL \\ Libraries}}}  \\
         \midrule
         LRP Toolbox~\cite{lrp_toolbox} & 2016 & LRP & Caffe\\
         \midrule
         DeepExplain~\cite{ancona2017towards} & 2017 & \makecell{Saliency maps, Grad * Input, $\epsilon$-LRP, DeepLIFT, DeConvNet} & Tensorflow, Keras\\
         \midrule
         iNNvestigate~\cite{JMLR:v20:18-540} & 2019 & \makecell{SmoothGrad, DeConvNet, Guided-BackProp, PatternNet, LRP\\ Input * Gradients, DeepTaylor, PatternAttribution, IG} & Tensorflow, Keras\\
         \midrule
         TorchRay~\cite{fong2019understanding} & 2019 & \makecell{Gradient, Guided-BackProp, Grad-CAM, RISE} & PyTorch\\
         \midrule
         Captum~\cite{captum} & 2019 & \makecell{SmoothGrad, DeConvNet, Guided-BackProp, PatternNet, LRP \\  Input * Gradients, DeepLIFT, DeepTaylor, LIME, Kernel SHAP, IG } & PyTorch\\
    \bottomrule
    \end{tabular}%
    }
    \label{tab:software}
\end{table}

\clearpage

\subsection{Methods}
\label{appx:methods}

\begin{table*}[!h]
  \caption{Summary of the XAI methods categorized by interpretability method employed, image modality and dataset. The spaces marked with a "-" mean that explanation is only provided through text sentences.}
  \label{tab:methods_alltogether}
  \resizebox{0.9\textwidth}{!}{
  \begin{tabular}{lcccc}
    \toprule
    \textbf{Method} & \textbf{Year} & \textbf{Interpretability Method} & \textbf{Modality} & \textbf{Dataset}\\
    \midrule
    Zhang et al.~\cite{zhang2017mdnet} & 2017 & Attention-based & Microscopic Images & BCIDR \\
    Jing et al.~\cite{jing2017automatic} & 2017 & Attention-based & X-ray & IU Chest X-ray, PEIR Gross \\
    Rajpurkar et al.~\cite{rajpurkar2018deep} & 2018 & CAM & X-ray & ChestX-ray8 \\
    Wang et al.~\cite{wang2018tienet} & 2018 & Saliency maps & X-ray & IU Chest X-ray,ChestX-ray14 \\
    Gale et al.~\cite{gale2018producing} & 2018 & SmoothGrad & X-ray & Pelvic X-ray \\
    Li et al.~\cite{li2018hybrid} & 2018 & Text-based & X-ray & IU Chest X-ray, CX-CHR (private) \\
    Malhi et al.~\cite{malhi2019explaining} & 2019 & LIME & Endoscopy & Red Lesion Endoscopy \\
    Young et al.~\cite{Young_2019} & 2019 & KernelSHAP & Dermoscopy & HAM10000 \\
    Eitel et al.~\cite{eitel2019testing} & 2019 & Occlusion & MRI & ADNI Initiative \\
    Sayres et al.~\cite{sayres2019using} & 2019 & Integrated Gradients & Fundus Images & Private \\
    Barata et al.~\cite{barata2019deep} & 2019 & CAM & Dermoscopy & ISIC 2017, ISIC 2018 \\
    Tschandl et al.~\cite{tschandl2019diagnostic} & 2019 & CBIR & Dermoscopy &  EDRA, ISIC 2017, Private Dataset \\
    Sun et al.~\cite{Sun2019} & 2019 & Text-based & Mammography & Inbreast \\
    Lee et al.~\cite{lee2019generation} & 2019 & Saliency maps & Mammography & CBIS-DDSM \\
    Lamy et al.~\cite{lamy2019explainable} & 2019 & CBR & Mammography & \makecell{BCW, Mammographic Mass, Breast Cancer} \\
    Singh et al.~\cite{singh2019chest} & 2019 & Stacked LSTM & X-ray & IU Chest X-ray \\
    Yin et al.~\cite{yin2019automatic} & 2019 & t-SNE & X-ray & IU Chest X-ray \\
    Liu et al.~\cite{liu2019clinically} & 2019 & Attention maps & X-ray & IU Chest X-ray, MIMIC-CXR \\
    Windish et al.~\cite{windisch2020implementation} & 2020 & Grad-CAM & MRI & IXI, Gliobastoma \\
    Magesh et al.~\cite{magesh2020explainable} & 2020 & LIME & DaTscan & PPMI \\
    Lin et al.~\cite{lin2020covid} & 2020 & Guided Grad-CAM & X-ray & COVIDx \\ 
    Lopatina et al.~\cite{lopatina2020investigation} & 2020 & DeepLIFT & MRI & Private \\
    Graziani et al.~\cite{graziani2020concept} & 2020 & TCAV & Microscopic Images & Camelyon16, Camelyon17 \\
    Margeloiu et al.~\cite{margeloiu2020improving} & 2020 & Adversarial Training & Dermoscopy & HAM10000 \\
    Rio-Torto et al.~\cite{rio2020understanding} & 2020 & In-model & Microscopic Images & NIH-NCI Cervical Cancer \\
    Fang et al.~\cite{fang2020concept} & 2020 & Concept-based & Microscopic Images & Infectious Keratitis Dataset \\
    Chen et al.~\cite{chen2020generating} & 2020 & Multi-Head Attention & X-ray & IU Chest X-ray, MIMIC-CXR \\
    Silva et al.~\cite{silva2020interpretability} & 2020 & CBIR, Saliency Map & X-ray & CheXpert \\
    Punn et al.~\cite{punn2021automated} & 2021 & LIME & X-ray & COVID-19 Dataset \\
    Wang et al.~\cite{wang2021interpretability} & 2021 & SHAP & Dermoscopy & HAM10000 \\
    Billah and Javed~\cite{billah2022bayesian} & 2021 & BCNN & Microscopy Images & ALL-IDB \\
    Barata et al.~\cite{barata2021improving} & 2021 & CBIR & Dermoscopy & ISIC 2018 \\
    Thiagarajan et al.~\cite{thiagarajan2021explanation} & 2021 & t-SNE & Microscopic Images & Breast Histopathology \\
    Barnett et al.~\cite{barnett2021interpretable} & 2021 & Prototype Activation Map & Mammography & Mammography Private Dataset \\
    Kim et al.~\cite{kim2021interpretation} & 2021 & Counterfactual Examples & X-ray & Chest X-ray 14, VinDr-CXR \\
    Schutte et al.~\cite{schutte2021using} & 2021 & Grad-CAM & X-ray/Microscopy Images & Osteoarthritis X-ray, Camelyon16 \\
    Kim et al.~\cite{kim2021xprotonet} & 2021 & Saliency maps & X-ray & NIH chest X-ray14 \\
    Singh et al.~\cite{singh2021interpretable} & 2021 & Prototype Activation Maps & X-ray & CORD-19 \\
    {Liu et al. ~\cite{liu2021exploring}}	& 2021 & Text-based	& X-ray	& IU Chest X-ray, MIMIC-CXR \\
    {Liu et al.~\cite{liu2021ACL}} &	2021 & Text-based &	X-ray &	IU Chest X-ray, MIMIC-CXR \\
    Lucieri et al.~\cite{lucieri2022exaid} & 2022 & TCAV & Dermoscopy & ISIC 2019, Derm7pt, PH2, SkinL2 \\
    Hu et al.~\cite{hu2022x} & 2022 & CBIR & X-ray/Dermoscopy & COVIDx, ISIC 2019 \\
    Gour and Jain~\cite{gour2022uncertainty} & 2022 & Uncertainty-based & X-ray & COVID19CXr \\
    {Yuksekgonul et al.~\cite{yuksekgonul2022post}} &	2022 &	CBM &	Dermoscopy &	HAM 10000, ISIC 2020 \\
    {Wang et al.~\cite{wang2022medical}} &	2022 &	Text-based &	X-ray &	MIMIC-CXR \\
    Singla et al.~\cite{singla2021explaining} & 2023 & Counterfactual Examples & X-ray & MIMIC-CXR \\
    {Yan et al.~\cite{yan2023towards}} &	2023 &	CBM & Dermoscopy &	ISIC 2016-2020 \\
    {Selivanov et al.~\cite{selivanov2023medical}}	& 2023 &	Text-based &	X-ray &	IU Chest X-ray, MIMIC-CXR \\
    {Zhang et al.~\cite{zhang2023novel}} &	2023	& Text-based &	X-ray &	IU Chest X-ray, MIMIC-CXR, COV-CTR \\
    {Wang et al.~\cite{wang2023metransformer}}	& 2023 &	Text-based &	X-ray &	IU Chest X-ray, MIMIC-CXR \\
    {Chen et al.~\cite{chen2023visual}} &	2023	& Text-based &	X-ray &	IU Chest X-ray, MIMIC-CXR \\
    {Yang et al.~\cite{yang2023radiology}} 	& 2023 &	Text-based &	X-ray &	IU Chest X-ray, MIMIC-CXR \\
    {Patrício et al.~\cite{patricio2023CVPRW}}	& 2023 &	CBM &	Dermoscopy &	PH2, Derm7pt \\
    \bottomrule
  \end{tabular}
  }
\end{table*}

\end{document}